\documentclass[5p]{elsarticle}
\usepackage{float}
\usepackage{graphicx}
\usepackage{hyperref}
\usepackage{cleveref}
\usepackage{booktabs}
\usepackage[tight]{subfigure}
\usepackage{caption}
\journal{Information and Software Technology}

\begin{document}

\begin{frontmatter}

%\title{Preventing Technical Debt by Technical Debt Aware Project Management}
\title{Preventing technical debt with the TAP framework \\ for Technical Debt Aware Management}
%\tnotetext[mytitlenote]{Fully documented templates are available in the elsarticle package on \href{http://www.ctan.org/tex-archive/macros/latex/contrib/elsarticle}{CTAN}.}

%% Group authors per affiliation:
%%\author{Marion Wiese}% \fnref{myfootnote}}
%%\address{Vogt-Kölln-Str.30, Hamburg}
%\fntext[myfootnote]{Since 1880.}

%Idee / passt das: Findings oder Contributions durch Boxen hervorheben?

%% or include affiliations in footnotes:
\author[mymainaddress]{Marion Wiese\corref{mycorrespondingauthor}}
\cortext[mycorrespondingauthor]{Corresponding author}
\ead{marion.wiese@uni-hamburg.de}

%\author[mymainaddress,mysecondaryaddress]{Paula Rachow}

\author[mymainaddress]{Paula Rachow}
\ead{paula.rachow@uni-hamburg.de}

\author[mymainaddress]{Matthias Riebisch}
\ead{matthias.riebisch@uni-hamburg.de}

\author[gujmainaddress]{Julian Schwarze}
\ead{schwarze.julian@guj.de}

\address[mymainaddress]{Universität Hamburg, Department of Informatics, Vogt-Kölln-Str.30, 22527 Hamburg}
%\address[mysecondaryaddress]{\ead[url]{https://www.inf.uni-hamburg.de/}}

\address[gujmainaddress]{Gruner + Jahr GmbH, Information Technology, Baumwall 11, 22459 Hamburg}
%\address[gujsecondaryaddress]{https://www.guj.de/}

\begin{abstract}
% 231 Worte laut MS Word / 300 Worte sind erlaubt

\textbf{Context.} 
%Technical Debts (TD) are problems of the internal software quality. 
%They are a metaphor for what is known as quick fixes and workarounds in practice.
Technical Debt (TD) is a metaphor for technical problems that are not visible to users and customers but hinder developers in their work, making future changes more difficult.
TD is often incurred due to tight project deadlines and can make future changes more costly or impossible.
Project Management usually focuses on customer benefits and pays less attention to their IT systems' internal quality.
%Their awareness of TD is usually low.
TD prevention should be preferred over TD repayment because subsequent refactoring and re-engineering are expensive. %more expensive than building the right solution from the beginning.
%While there are numerous works on TD repayment, solutions for TD prevention are understudied.

\textbf{Objective.} 
This paper evaluates a framework focusing on both TD prevention and TD repayment in the context of agile-managed projects. 
The framework was developed and applied in an IT unit of a publishing house. 
The unique contribution of this framework is the integration of TD management into project management. 

\textbf{Method.}
The evaluation was performed as a comparative case study based on ticket statistics and two structured surveys. 
The surveys were conducted in the observed IT unit using the framework and a comparison unit not using the framework. 
The first survey targeted team members, the second one IT managers.

\textbf{Results.} 
The evaluation shows that in this IT unit the TAP framework led to a raised awareness for the incurrence of TD. 
Decisions to incur TD are intentional, and TD is repaid timelier. 
Unintentional TD incurred by unconscious decisions is prevented. 
Furthermore, better communication and better planning of the project pipeline can be observed.

\textbf{Conclusions.}
We provide an insight into practitioners' ways to identify, monitor, prevent and repay TD.
The presented framework includes a feasible method for TD prevention despite tight timelines by making TD repayment part of project management.

\end{abstract}


\end{frontmatter}

%\linenumbers

	\section{INTRODUCTION} 
	\label{section:introduction}
% ## Motivation ## 
% Software systems are subject to frequent changes during development and usage, mostly due to changes of needs and environment. 
% Developers experience growing difficulties. Number of errors increases. 
% Noticed: effort for further changes raises continuously. 
% discovered reasons: introduced problems, some of them consciously due to time pressure. 
% To be repaired by refactoring, however often delayed. 
% increase of effort similar to a penalty: name technical debt coined as a metaphor. 
% reduction of technical debt is important for evolution, and its prevention is even better.
    Software systems are subject to frequent changes due to new or changing requirements or changes in the environment. 
    Hence, a central goal in the development of software systems should be to implement these systems in such a way that they are easily changeable over a long time. 
    Many different causes limit this ease of changeability and should be addressed. 
    Because these modifications usually provide no visible benefit to non-developers and are challenging to understand, they are often given a lower priority by business leaders.
    Cunningham introduced the metaphor of Technical Debt (TD) to make this problem better understandable to business managers \cite{Cunningham1992}.

% ## Begriffseinführung ## 
    Avgeriou et al. define TD as ``\textit{a collection of design or implementation constructs that are expedient in the short term, but set up a technical context that can make future changes more costly or impossible}'' \cite{Avgeriou2016a}.
    In a technical metaphor for financial debt, a sub-optimal implementation or design is interpreted as debt. 
	The resulting problems are interpreted as interest rates, refactoring as repayment, and refactoring cost as principal.
	%TD is a serious problem in practice and often caused by tight deadline due to fast changing requirements. %and much research has been done on this topic. 
	%In 2016 Avgeriou et al. claimed: ``\textit{It’s the next big thing, and it’s messy.}'' \cite{Avgeriou2016d}.
	
    %Avgeriou et al. \cite{Avgeriou2016a} describe this as follows: ``\textit{While the conceptual roots of technical debt imply an idealized, deliberate decision-making process and rework strategy as needed, we now understand that technical debt is often incurred unintentionally and catches software developers by surprise}''.
    In 2016, Avgeriou et al. claimed about TD: ``\textit {It's the next big thing, and it's messy}''\cite{Avgeriou2016d}.
	TD is a serious problem in practice, often caused by tight deadlines due to fast-changing requirements in the age of digitalization \cite{Rios2020, Avgeriou2016a, Verdecchia2021a}. %and much research has been done on this topic. 
	
    McConnell and Fowler differentiate between intentional and unintentional incurred TD \cite{McConnell2008a,Fowler2019}. 
	Intentional TDs result from a conscious decision, e.g., implementing a work\-around to reach a set deadline.
	Unintentional TDs result from mistakes, e.g., poor design at the code level or architectural decisions that, over time, turn out to have been wrong.
	
% ## Repayment vs. Prevention ## 
    \textbf{\textit{TD prevention}} is stated as the preferable option for TD management by many practitioners \cite{Tsoukalas2020, Freire2020a, Apa2020}. 
   % Practitioners often state TD prevention is the preferable option for TD management \cite{Tsoukalas2020, Freire2020a, Apa2020}.  %PreventionBeforeRepaymentPractitioners
    In a long-term view, implementing the optimal solution right from the beginning is cheaper than incurring TD, especially when considering possible interest payments \cite{Yli-Huumo2016, Rios2020}. %PreventionBeforeRepayment
	% When asked, practitioners often state that mechanisms for TD prevention should be prioritized before mechanisms for repayment \cite{}.
%	Therefore, this work focuses on the prevention part of the framework. 
	TD prevention primarily targets unintentional TD.
	To enable TD prevention, two main aspects must be considered: 
	\begin{enumerate}
	    \item Stakeholders have to be aware of TD to make conscious decisions \cite{Morgenthaler2012, Besker2020d, Perez2021a}. %Gupta2016,BOUGOUFFA201870, Moreira2021, Rios2018a,   %Awareness=>Prevention
	    \item Stakeholders have to understand the causes of TD \cite{Rios2020, Freire2021a}. %Causes=>Prevention %Pacheco2019, 
	\end{enumerate} 
	
% ## Overview => Awareness
    \textbf{ \textit{TD Awareness.}}
    Awareness is self-evident for self-admitted TD, i.e., ToDo-Statements in the code, and TD found by tools. % on three levels of situational awareness. 
    In practice, TD prevention is often focused on these kinds of TD \cite{Rocha2017, Apa2020}. %code debt + PreventionSolution % Sneed2014
    %In contrast, the framework presented in this paper raises the awareness for design and architecture decisions and the TD incurred by these.
    In research, the perception of design and architecture decisions and the TD incurred by these is less evident and less focused on.
    
    TD backlogs and lists lead to more visibility, make TD explicit, and raise the perception of TD.
    This, in turn, raises the overall awareness for TD \cite{Kruchten2012a, Eliasson2015, Guo2016a, Besker2019c}.
    %The backlog resulting from the adoption of this framework makes TD visible and explicit which enhances the overall awareness for TD \cite{Kruchten2012a, Eliasson2015, Guo2016a, Besker2019c}. %Treude2010,  %Overview=>Awareness
    
% ## Awareness => Communication %CommunicationProblemStatement
    The awareness leads to better communication between developers and IT management, which is a problem often mentioned regarding TD management \cite{Avgeriou2016d, Soliman2021a, Borowa2021, Freire2021a}. 
    %Lim2012, Martini2015a
    For example, in their work about architectural TD (ATD), Verdeccia et al. found ``\textit{difficulties in communicating the presence of ATD to the stakeholders of a software product}''\cite{Verdecchia2021a}.
    
% ## Awareness => Pipeline/Planning    %ProjectPlanningProblem
    %Better communication, in turn, can effect the planning of the project pipeline by the IT managers, i.e., the ordering of different projects of the project portfolio by time.
    Better communication and the overview of TD can, in turn, impact how IT managers plan the project pipeline, i.e., the timing of the project portfolio's various projects.
    This could provide a solution to the impediment mentioned by Besker et al. \cite{Besker2018b}:  
    ``\textit{If the managers are not aware of the amount of software development time the developers waste because of TD, they are consequently not able to react and take appropriate action regarding the wasted time.}''

% ## Causes allgemein (inkl. Timeline Problem) ## 
    %Additional to raising Awareness, one has to understand the causes to enable TD prevention \cite{Pacheco2019, Rios2020, Freire2021a}. %Causes=>Prevention
    \textbf{ \textit{TD Causes.}}
    The initially named cause for TD incurrence is tight project deadlines, as described in the paper of Cunningham \cite{Cunningham1992}. 
    Other causes have been identified by different research papers, e.g., bad design decisions, unavailability of a key person, neglected technical improvements, lack of education, communication issues, or problems related to project planning \cite{Kruchten2012a, Martini2014b, Li2015, Rios2020}. %TD Causes %Lenarduzzi2019d, Bi2021a,
	% Gekürzt: Ernst2015b, Yli-Huumo2016,  Avgeriou2016a, Verdecchia2021a, Perez2021a, Borowa2021, Soliman2021a
% ## Tight Timeline Problem ## 
    Nevertheless, tight deadlines are identified by many research works as the most pressing cause \cite{Martini2014b, Ernst2015, Rios2020, Verdecchia2021a, Freire2021a, Soliman2021a}. %TightTimelineProblemStatement
    % Moreira2021, Peters2014a, Bi2021a, Lim2012, Avgeriou2016a, Guo2016d, Ramac2020b, Freire2020a, Rocha2017, Tsoukalas2020, Lenarduzzi2019g, Besker2019c, 

% ## Beispiel (Timeline-bezug) ##
    An example of the problem of tight deadlines is the German government's VAT cut in the summer of 2020 as part of its response to the Corona crisis. 
    The corresponding law was enacted only two days before it came into force. 
    To reflect this change in the software systems, adjustments to these systems were necessary for many German IT units.
    These units were, therefore, subject to a tight schedule. 
    %Well-structured project planning or adoption of good development practices were mostly not possible. 
   
% ## Developer Phenomenon / Decision-Making## 
    In many cases, this constant time pressure leads to a phenomenon in developer behavior. 
    In preemptive obedience, the developers search for the fastest and easiest solution assuming a tight schedule, even when there is no time pressure. 
    To step back and think about or discuss different solution options is not common practice.
    In a paper about cognitive biases and their antecedents, Borowa et al. call this phenomenon communicational antecedents, i.e., ``\textit{constructive criticism and challenging the ideas underlying the decisions of others is not standard}''\cite{Borowa2021}.
    %This may lead to the anchoring bias which means to choose the first available option.
    This may lead to the anchoring bias, i.e., choosing the first available option.
    In turn, this bias causes decisions to incur TDs that are not made intentionally. 
    Regarding this topic, Besker et al. found that ``\textit{practitioners rarely base their decisions on pre-defined procedures or guidelines}''\cite{Besker2019c}, and Lenarduzzi et al. state that ``\textit{decisions related to TD issues were often informal and ad-hoc}''\cite{Lenarduzzi2019g}.
    
    There may be frequent situations where the sub-optimal but faster to implement solution must be chosen due to project deadlines.
    However, it should be a goal to make it the whole team's conscious and intentional decision to choose the optimal or sub-optimal solution.
    
% ## Framework##  
    \textbf{\textit{In this paper}}, we present and evaluate the TAP framework for \textbf{T}echnical debt \textbf{A}ware \textbf{P}roject management which was developed by practitioners.
	The TAP framework divides all tasks that are not visible to the customer into four categories.
	The developers create corresponding backlog tickets.
    One of these categories is the so-called TD ticket category.
    These tickets comprise only intentional TD and relate these tickets to the project incurring the TD.
    The repayment of these TD items is performed as part of the project after meeting the set deadline.
    Other parts of the TAP framework are a continuous time contingent for maintenance and maintenance projects similar to what McConnell described \cite{McConnell2008a}.   
    Maintenance in the TAP framework includes repayment of unintentional TD, adaption to new technologies, optimization of quality attributes. 
    This way of handling TD is a new contribution to the State of the Art.
    
% ## TD tickets ##  

    To the best of the authors' notice, no studies integrate TD management into the project management process focusing on the time limit and including a solution to prevent TD despite tight deadlines that is feasible in an industrial environment.
    By  TD tickets and their effects on decision-making and the project pipeline, this framework presents such a solution.

% ## Contribution ##   
    %\textbf{Contribution.} 
    
	\textbf{\textit{The contribution}} of this paper is the presentation and evaluation of the TAP framework and especially the TD tickets.
	The evaluation consists of the analysis of ticket statistics and two surveys addressing team members and IT managers, respectively. 
	
	The evaluation will show the TAP framework's benefits as perceived by the IT unit using it:
	\begin{itemize}
    \setlength{\parskip}{1pt}
	    \item Communication between different stakeholders is optimized.
	    \item Overall awareness for TD is raised.
	    %\item Project managers get an extrinsic motivation to reduce the incurrence of TD.
	    \item Decisions on TD incurrence are made consciously and intentionally.
	    \item TD incurred by unintentional and unconscious decisions can be prevented.
	    \item TD incurred by intentional and conscious decisions due to tight deadlines are repaid timely.
	\end{itemize}
	Additionally, a survey of the IT managers indicates that they can plan the project pipeline more realistically and make the customer understand delays due to TD repayment and prevention.
		    
% ## Outline ##   
	In the following subsection, we will give an overview of the related work. 
	Subsequently, we will motivate and introduce the TAP framework in \Cref{section:framework}. 
	\Cref{section:csmethod} will present the case study design, including research questions (RQs) regarding the TAP framework's effects, benefits, and drawbacks. 
	The evaluation comprises ticket statistics and results of two surveys conducted in the observed IT unit and a comparison unit. 
	The evaluation's results will be presented in \Cref{section:results}.
	In \Cref{section:discussion}, we will discuss the results, answer the RQs, and describe the threats to validity.
	The paper ends with the conclusion and future work in \Cref{section:conclusion}.

	\section{RELATED WORK}
	\label{section:relatedwork}
    In this section, we present the related work to our paper. 
    For this purpose, we follow the paper's title and describe the state of the art of TD prevention, TD awareness, TD in project management, and case studies for TD management frameworks or methods.
        \subsection{TD Prevention}
        %``TD prevention aims to prevent potential TD from being incurred.`` \cite{Li2015}
        TD Prevention is often mentioned in research but mostly as a \textbf{problem statement} \cite{Freire2020, Tsoukalas2020, Vogel-Heuser2021}. %PReventionProbelmStatement
        %He2016,  Stopford2017, Fontana2017, Ampatzoglou2016a, Digkas2020, Rios2021, Rios2018
        In 2015, Li et al. \cite{Li2015} found TD prevention to be understudied.
        More research on how to prevent TD beyond raising TD awareness has been done since.
        
        Many papers researching \textbf{prevention strategies} focus on inadvertent code debt and tools helping to find these debts, e.g., code smells, before incurrence \cite{Yli-Huumo2016, Apa2020, Tsoukalas2020}. %code debt + PreventionTopic (+ PreventionSolution)
        %Digkas2020, Sneed2014, Freire2020b, Rocha2017, 
        %Papers focusing on the prevention of TD on design and architectural level which is mostly incurred during decision-making are fewer. 
        Fewer papers focus on the prevention of TD on the design and architecture level, which are usually incurred during decision-making.
        
        Morgenthaler et al. provided insights to the TD management of build debts at Google. They summarized their prevention strategies in three topics ``Automation'', ``Make it easy to do the right thing'', and ``Make it hard to do the wrong thing'' \cite{Morgenthaler2012}.
        Yli-Huumo et al. found that minimum viable products can help prevent TD by focusing TD prevention on areas that are important for the customer and throwing away parts that have shown no real business value \cite{Yli-Huumo2015}. 
        Preventing documentation debt during requirements engineering was the research focus of Charalampidou et al. 
        They integrated requirements specifications into the IDE (integrated development environment) \cite{Charalampidou2018}.
        %Regarding TD prevention, 
        Lenarduzzi et al. suggested optimizing requirements and receiving early feedback from customers. 
        These approaches improved the effort estimation and enhanced the quality by optimizing tests and documentation \cite{Lenarduzzi2019g}.
        Yang et al. suggested for COTS (commercial of the shelf) - intensive systems to train the COTS assessment team and explore contracting options with COTS vendors \cite{Yang2019c}. %Streichkandidat
        Another way to prevent TD is to reuse code which was mentioned in the work of Feitosa et al. \cite{Feitosa2020}. 
        In their comparison of encouraging, rewarding, forcing, and penalizing strategies for TD reduction, Besker et al. found some companies were using educational sessions on how to avoid TD \cite{Besker2020d}.
        Vogel-Heuser et al. advised using guidelines to prevent TD in embedded systems\cite{Vogel-Heuser2021}.
        Borowa et al. presented some mechanisms to prevent cognitive biases that may lead to TD incurrence, such as ``challenging all decisions'', ``explicitly gathering information about alternatives'', ``documenting and passing on knowledge'', ``explicitly registering all accounts of TD'' and ``defining and recording who is responsible'' \cite{Borowa2021}.
        
        The InsighTD project\footnote{\url{http://www.td-survey.com/}} is an international survey study.
        The survey participants often stated that their TD could have been prevented \cite{Rios2020}.  %InsighTD + PreventionProblemStatement %Rios2018, Rios2021
        Various papers of this study uncovered TD causes and prevention strategies used by practitioners.
        The most commonly found prevention strategies are ``well-defined scope/requirements'', ``code evaluation/standardization'', ``following well-defined project processes'', ``well-defined architecture/design'', ``TD awareness/management'', ``adoption of good practices'', ``improving tests/coverage'', ``good communication'', and ``reviews'' \cite{Rios2020, Freire2020, Freire2020a, Perez2021a}. %InsighTD + PreventionSolution 
        %folgende Paper werden sonst nicht zitiert: Rios2019d, Moreira2021(nur in Sammlungen)
        
        All these research results are valid aspects of TD prevention. 
        However, compared to the \textbf{TAP framework} these aspects are mostly vague, i.e., details and information on how to achieve these aspects are missing. 
        Furthermore, they do not provide a solution to the problem of tight deadlines.

        \subsection{TD Awareness}
        In 2012, Kruchten et al. already mentioned the importance of TD awareness: ``\textit{The first step is awareness: identifying debt and its causes}'' \cite{Kruchten2012a}. 
 		In a survey with more than 1,800 participants, Ernst et al. found that 79\% of the participants strongly agree that ``\textit{Lack of awareness of TD is a problem}'' \cite{Ernst2015}. 
        
        As with prevention, there are many papers mentioning TD awareness in the form of a \textbf{problem statement} \cite{Morgenthaler2012, Sharma2015, Apa2020, Vogel-Heuser2021}. %AwarenessProblemStatement 
        %Martini2015a, Besker2017d, Besker2017c, Vogel-Heuser2017, Besker2018b, Stopford2017, Klotins2018a, Cico2019, Mandic2020a, Fontana2017, Guo2011b,   Martini2016b, Holvitie2018b, Tom2013b, Moreira2021, Besker2019c,  Ernst2015, Baysal2013, Treude2010, 
        
        %Overview=>Awareness
        Papers on \textbf{how to raise awareness} are less common.
        Many research papers found that visualization helps to raise awareness for TD.
        An obvious way to visualize TD is to create a list of all known TD items, which is also part of the TAP framework.
        For example, Kruchten et al. provided a color scheme for a backlog including colors for new features, architectural features, defects, and TD \cite{Kruchten2012a}. %Overview=>Awareness
        Kaiser et al. presented the Code Christmas Tree, which is a visual presentation of the code coverage of a system. 
        This visualization was hung in the hallway and resulted in discussions.
        By this, the visualization raised the awareness of different stakeholders \cite{Kaiser2011}.
        Gupta et al. also found that so-called information radiators, i.e., visualizations that are displayed where people can see it frequently, were a good way to raise awareness \cite{Gupta2016}.
        Baysal et al. as well as  Treude et al. examined the positive effects dashboards have on developers' awareness of their projects' issues \cite{Baysal2013, Treude2010}.
        Guo et al. presented a TD management approach including a list of TD, which increased the awareness of TD. 
        In this case study, the project manager stressed the TD list's positive effect \cite{Guo2016a}.
        Eliasson et al. proposed a method to visualize dependencies in the automotive domain to make TD visible. 
        They found their method to be useful for stakeholders to be aware of architectural TD \cite{Eliasson2015}.
        Martini et al. proposed a method called AnaConDebt to identify whether and when Architecture TD should be refactored.
        They found this method is useful to raise awareness ``\textit{about the importance and the urgency of proactively refactoring ATD}'' \cite{Martini2016c}.
        Baars et al. used a gamification approach to make developers aware of harmful patterns in their codebase by turning the patterns into monsters in Minecraft\footnote{\url{https://www.minecraft.net/}} \cite{Baars2019a}.
        Al Mamun et al. researched TD in self-driving miniature cars.
        They focused on code debt and proposed that tools detecting rules violations would also increase awareness \cite{AlMamun2014}.
        
        %Communication/Training=>Awareness
        Another way to raise awareness is to optimize communication, e.g., workshops and training sessions. The communication aspect is important for the TAP framework, too.
        Shrama et al. found the awareness of the refactoring's benefits was lacking. They reported on the strategy to conduct workshops on the refactoring's benefits and refactoring tools \cite{Sharma2015}.
        Letouzey et al. also mentioned an awareness workshop for top managers to introduce TD and the SQUALE\footnote{\url{http://www.squale.org/}} method to them \cite{Letouzey2016}.
        Tonin et al. \cite{Tonin2017} researched the effects of TD awareness by conducting a classroom study. 
 		The students were informed about the concept of TD and should use TD boards during a project. 
		The effects of this were a changed attitude, more discussions, and more conscious incurrence of TD. 
       
        In addition to the presented ways to raise TD awareness, the \textbf{TAP framework} raises the awareness by assigning responsibilities for TD management which is an aspect not found in research to the best of the authors' notice.
      
        \subsection{TD management and project management}
        %, but is the main focus of our work and solved by integrating TD management into project management.
        A project is characterized by a beginning and an end and thus has a deadline. 
        This deadline is an often mentioned cause for TD, as shown in \Cref{section:introduction}.
        One often mentioned \textbf{problem} for managing and preventing TD is related to project management \cite{Guo2016a, Rios2020, Freire2020a, Perez2021a}.
        ``\textit{Lack of process integration}'' is an obstacle mentioned in the work of Guo et al.
        Their case study provided some insights into the obstacles of adapting research methods to a practical environment \cite{Guo2016d}.
        Martini et al. found that the ``\textit{split of budget in project budget and maintenance budget boosts the accumulation of debt}'' \cite{Martini2014b}. 
        
        To \textbf{solve these problems} Ramasubbu et al. integrated TD management into quality management steps \cite{Ramasubbu2019a}. 
        They created a TD register to store TD items in relation to their requirements and resulting defects. 
        Furthermore, they considered design moves to implement requirements or repay TD and the risks associated with them. 
        They conducted a cost-benefit analysis on this data and proposed a way to control TD. % with these information. 
        While their approach is similar to the TAP framework, their focus was on companies that already use quality assurance processes, and their approach did not focus on TD prevention. 
        Other research papers address the management of a TD backlog, e.g. \cite{Schmid2013a, Kruchten2012a, Bachmann2012}. 
        
        In contrast to the \textbf{TAP framework}, these papers do not focus on projects and the resulting timeline problems.
        The TAP framework tackles the deadline problem by integrating TD management into project management, i.e., integrating the repayment of a project's TD as part of the project itself.
        
        \subsection{Case Studies on TD Management}
        Regarding TD management, Besker et al. found that it ``\textit{promotes a culture within organizations in which developers are supported and appreciated for managing their TD}'' and that it ``\textit{leads to a virtuous cycle where the right culture and TD prevention mechanisms reinforce each other}'' \cite{Besker2020c}.
        
        This paper presents a comparative case study of a TD management framework developed by practitioners. 
        To the best of the authors' notice, no other case studies compare an IT unit using a framework or method for managing TD with an IT unit not using a TD management strategy.  
        For this reason, we compare our work to other case studies on \textbf{TD management frameworks and methods} but focus on confirmatory case studies and action researches. 
    
        %Exploratory case studies have been done use data, e.g. interview, survey or ticket data  from case companies or teams to get to know the state of the art of TD Management.
        %For example Yli-Huumo et al. studied eight team in a company an developed a maturity model for each TD activity from this data \cite{Yli-Huumo2016}.
        In an action research, Yli-Huumo et al. developed a process for identification, documentation, and prioritization of TD.
        They focused on uncovering the risks and benefits related to these TD activities \cite{Yli-Huumo2016b}.
        Guo et al. focused their case study on uncovering the costs of their TD management approach, which centers around a TD list and comprises five steps for measuring and prioritizing TD items. 
        During the implementation of this approach in practice, they had to face various obstacles, which they presented in another paper \cite{Guo2016d}.
        Finally, they applied their approach not directly in practice but simulated it for past releases on actual company data. 
        %They found the initial costs in the first sprint to be higher than the following sprints but still lower than an average use case. 
        %``Analysis and evalutaion'' was the activity with the most costs \cite{Guo2016a}.
        The Tracy framework for business-driven TD prioritization was developed by Reboucas de Almeida et al.
        They proposed this framework to prioritize TD repayment activities not only by technical necessities but also by the business value of the configuration item that holds the TD item \cite{DeAlmeida2021}. 
        %In their works, they conducted an exploratory case study, developed the framework, and finally evaluated their approach in an industrial case study \cite{ReboucasDeAlmeida2018c, DeAlmeida2019, DeAlmeida2021}.%Streichkandidat
        %Explorative case studies to understand the TD concepts are often done \cite{}.
        %Confirmative case studies to test comprehensive TD management frameworks in a practitioners environment are fewer.
        The work of Ramasubbu et al. is similar to the TAP framework and was also validated by an industrial case study \cite{Ramasubbu2019a}. 
        We already introduced it in the previous paragraph.
        Malakuti et al. described the TD management approach of a specific project and obstacles while introducing TD concepts to a company, but they were still missing a working solution \cite{Malakuti2020}.
        
        All these TD management strategies focus on the general management of TD without a relation to a specific time frame.
        In contrast, the central aspect of the \textbf{TAP framework} 's TD tickets is the integration of TD management into project management which includes the deadline aspect.

	\section{TAP FRAMEWORK}
	\label{section:framework}
	
    	\subsection  {Motivation}
    	\label{section:frameworkMotivation}
    % Basics
    	The TAP framework was developed and established in an IT unit at the beginning of 2018. 
    	It was utilized ever since and is the subject of this case study. 	
    	The focus of the IT unit that developed this framework is on systems that support the marketing of advertisements in print, online and mobile media, a volatile and competitive market.
    
    % TD Causes
    	Accordingly, tight timelines and frequently changing systems are the main cause for the TD of this IT unit.
    	%There is a high market pressure in this industry sector and contracts or laws need to be fulfilled in time. 
    	The tight timelines often meant that ``\textit{well-structured project planning}'' or ``\textit{adopting good development practices}'' as proposed by Freire et al. \cite{Freire2020} is not sufficient.
    	
    % TD consequences	
    	In this IT unit, TD accumulation led to increasing cycle times, unnecessary errors in the systems, and developers' discontent.
    	The IT management was surprised by these problems because they were not aware of having incurred debts in any form.
    % Call to Action	
    	Ultimately, however, they recognized these problems and gave sufficient priority to dealing with TD. 
        The IT unit including the management was willing to change its processes.
    
    % Action Goal    
        The main goals for this change were pre-defined by the IT unit's manager: 
        First, TD must be prevented in any case if this is possible without affecting the deadlines. 
        Second, whenever a set deadline leads to TD, this TD shall be repaid timely. 
        Third, the IT managers should have the ability to track the TD in their systems.
        %surprise.
        
    % Result
        On this basis, a framework to help manage TD was established. 
        Two factors of the TAP framework, in particular, were intended to support TD prevention: 
        the overall raised awareness for TD incurrence and the integration of TD management into the project management process.
        A third factor, the overview of TD resulting from the framework adoption, helped the management track TD.
        The resulting TAP framework for TD management is presented below.
		
		\subsection  {Development}
		\label{section:FrameworkDevelopment}
		The framework was developed in a working group of two managers and two solution architects of the IT unit. 
		The development spanned five steps: $(I)$ identification of the tasks that hinder the development of functional requirements, $(II)$ identification of management and prevention processes for these tasks, i.e., ticket categories, $(III)$ assignment of tasks to ticket categories, $(IV)$ development and documentation of the overall framework, $(V)$ establishment of the new processes in the IT unit.
		%The establishment spanned four more steps: $(a)$ introduction to the unit and $(b)$ to the sub units by the managers and $(c)$ introduction to the Scrum teams and $(d)$ support and guidance by the architects.
		
		Eleven hindering tasks were identified: $(1)$ bug fixing, $(2)$ regularly refactoring of code that may be in a bad condition due to causes that are not specified in detail, $(3)$ workarounds which are made under the pressure of a deadline, $(4)$ missing documentation or tests mostly due to time pressure, $(5)$ implementation that stay incomplete because of changing business decisions, $(6)$ temporary switches in the program for the distinction of new and old program parts, $(7)$ adoption of new architecture approaches, $(8)$ adapting new infrastructure like e.g. version upgrades, $(9)$ innovation like proofs of concepts (POC) for new technologies, $(10)$ implementation and refinement of technical monitoring tools, $(11)$ performance optimization.
		
		Four processing types were developed to handle these tasks, resulting in the corresponding ticket categories. 
		In this framework, all other tasks are functional requirements including bugs and performance issues that are visible to the user. 
		The assignment of the hindering tasks to these categories can be seen in \Cref{tab:TaskCategories}.

		\begin{table}[h]
			\centering
			\begin{tabular}[h]{lll}
			\hline
			Task & Scope &  Category  \\ 
			\hline 
			$(1)$ Bug fixing & not visible/ & MT  \\ 
		 & visible & FRT  \\ 
			$(2)$ Refactoring & small & MT   \\ 
		 & big & MP \\ 
			$(3)$ Workaround &  & TDT   \\
			$(4)$ Documentation/Test &  & TDT \\ 
			$(5)$ Incomplete implementation & &  TDT \\
			$(6)$ Temporary switch &  & DT \\ 
			$(7)$ Architecture change & small & MT  \\ 
		 & big & MP \\ 
			$(8)$ Infrastructure &  & MP  \\ 
			$(9)$ POCs & small & MT  \\ 
		 & big & MP  \\ 
			$(10)$ Monitoring &  & MT\\ 
			$(11)$ Performance & not visible & MT \\ 
		 & visible & FRT  \\ 
			\hline
			\end{tabular}
			\captionsetup{justification=centering}
		\caption{Hindering tasks and their assigned ticket categories \\
		(MT - Maintenance Ticket, MP - Maintenance Project,\\
		TDT - TD Tickets, DT - Deconstruction Tickets,\\
		FRT - Functional Requirement Ticket)}
		\label{tab:TaskCategories}
		\end{table}
		
		\subsection  {Ticket Categories}
		\label{section:TicketCategories}
	
	    The TAP framework results in four ticket categories for handling different types of TD-related tasks.
	    It is essential to notice that the distinction of these categories is not because of the causes or theoretical classification schemes. 
	    The distinction refers to the way the tickets are to be processed.
	    A fifth ticket category handles all functional requirements.
	    The latter category is not presented in detail as it is not in the scope of this framework. 
	    
	    All tasks are recorded as tickets in a comprehensive project backlog and tagged and handled depending on their category.
		The framework includes guidelines for the category-specific recording and processing of these tickets.
		The person who has the greatest interest in completing the tasks of a category is the one responsible for this category. 
		This person may be an architect or a business analyst. 
	    \Cref{tab:solution} shows an overview of the ticket categories, the responsible person, and the processing conditions.

		\begin{table*} 
	       % \footnotesize
		    \centering
			%\begin{tabular}{p{3cm} p{2,3 cm} p{2,3 cm} p{4,1cm} p{3,2cm}}
			\begin{tabular}{lllll}
				\toprule
				Ticket Category		& Responsible person		& Recorded By		& Processing Time						& Contingent\\ 
				\midrule 
			%	\midrule
				Maintenance  		 	& architect	 		& developer			& continually/architect decision 	& maintenance (10\%) \\ 
			%	\midrule 
	 			Maintenance Project 	& architect 		& architect			& management decision				& project\\ 
			%	\midrule 
				Technical Debt  	 	& business analyst	& all 				& after deadline, part of project	& project \\
			%	\midrule 
	 			Deconstruction 	 	    & architect 		& all				& as soon as possible				& project/functional req.\\ 
			%	\midrule 
				Functional Req.  & business analyst & business analyst & business analyst decision			& project/functional req.\\ 
	 			\bottomrule
			\end{tabular}
			\caption{Ticket categories, responsible persons, and time}
			\label{tab:solution}
		\end{table*}

		\subsubsection {Maintenance}
		In the TAP framework, maintenance tickets are technically driven tickets that have no direct impact on the user's perception of the system\footnote{We use the term ``maintenance'' as it is often used in agile organizations. 
        However, in research, standardization, and sometimes in classical organizations, the implementation of new or changing functional requirements is also referred to as adaptive maintenance.}. % \cite{ISO/IEC2006a}. }
        In particular, adapting systems technically, enhancing quality attributes, and repaying unintentional TD are parts of this category.
    	%, but is the core of its agile development. It is therefore not included in this category. On the other hand, 
    	
		\textit{Recording:} Every maintenance ticket must contain one sentence describing the ticket's impact. 
		The description must be understandable to business analysts and managers. 
		
		\textit{Processing:} Ten percent of every sprint's planned capacity is invested in maintenance tickets according to the architect's decision.

		\textit{Goal:} The goal of the maintenance tickets is to repay the unintentionally incurred TD continually and to allocate time for other maintenance tasks, e.g., technical adaption.

	    \textit{Example:} 
	    Examples of this category are the upgrade of a third-party library or the refactoring of poorly structured code. 
	
		\subsubsection  {Maintenance Project}
		All tasks belonging to the maintenance category but requiring more than five days of development time are treated  as maintenance projects. 
		This is based on the practice applied for functional requirements and business projects. %where small functional adaptions are part of continual work and bigger requirements will be implemented in the context of projects.

		\textit{Recording:} %Streichen?
		%Maintenance projects are handled like a business project. 
		%Therefore, 
		The architects prioritize maintenance projects and create a pipeline of these projects. 
		
		\textit{Processing:} This pipeline is integrated into the overall project pipeline as defined and prioritized by the IT managers.
		
		\textit{Goal:} Maintenance projects aim to allocate enough time to perform maintenance tasks with a long duration or maintenance tasks requiring more organization, e.g., when more than one team is involved. 

	    \textit{Example:} 
		An example is a version upgrade of a central database used by more than one application. 
		Another example is the change of the system's architecture, e.g., from a SOA to a microservice architecture. 
		%Streichen?
		%Both will take a lot of time and needs the collaboration of different teams as more than one application is affected. 
		%This in turn makes a project planning necessary.

		\subsubsection  {Technical Debt}
		These so-called TD tickets only comprise intentional TD. 
		These tickets describe tasks necessary to improve internal software quality during the implementation of a functional requirement, e.g., time for clean code, good design, or to comply with a specified architecture. 
		These tasks are not mandatory to implement the required functionality and can be implemented after a given deadline.% if this is necessary for the project.
		
		The team often discusses two or more solution options, one of which follows the standards and the specified architecture, and the others more or less deviate from the optimal solution.
		The latter, however, may be faster to implement. 
		Hence, this category only refers to conscious decisions to incur TD, i.e., intentional TD.
		
		This idea of TD follows the original introduction of TD by Cunningham \cite{Cunningham1992}. 

		\textit{Recording:} In such cases, the team consciously decides to record two tickets: 
		The functional requirement ticket describes the sub-optimal solution implemented before the deadline. With this ticket, the team incurs TD. 
		The TD ticket describes the optimal solution and the tasks to repay the incurred TD. 
		TD tickets are usually identified during estimation meetings when the team discusses the details of a functional requirement and estimates the effort for this requirement. 
		%Streichen?
		%This is also why, most of the TD tickets span mainly design and architectural TD and not code or other TD, e.g. unintentional TD.
		
\begin{figure} % [ht]
	 		\centering
			\includegraphics[width=\linewidth]{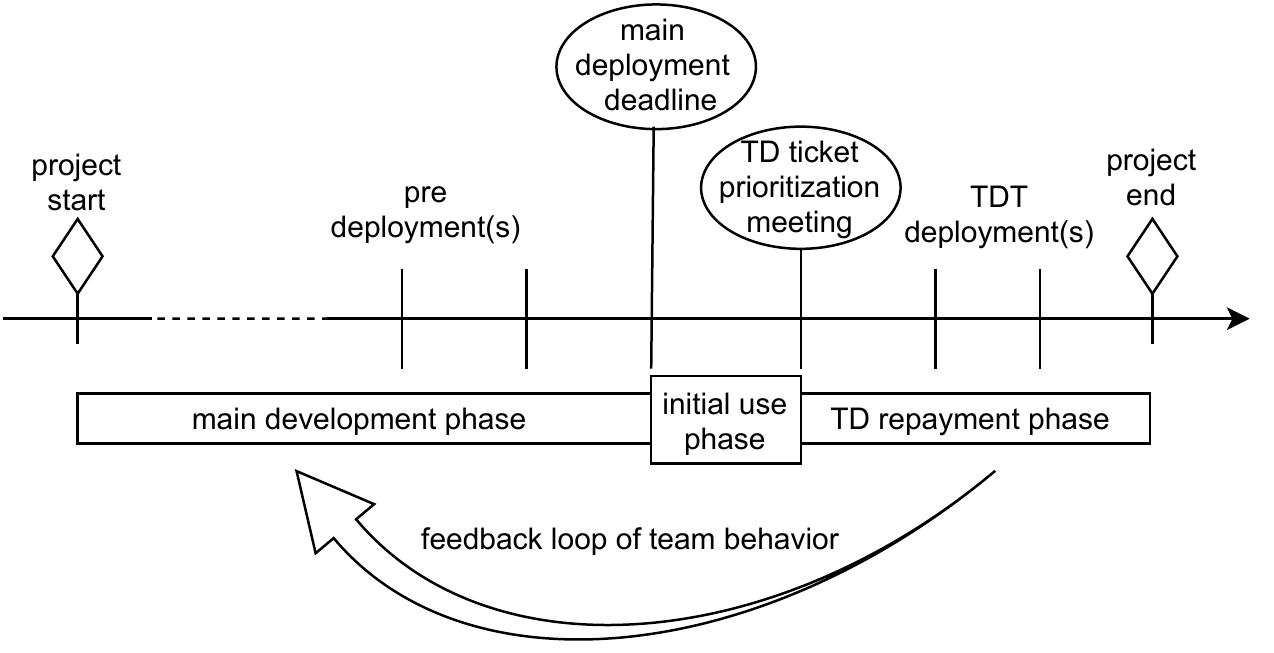}
			\caption{Project plan with included TD repayment phase}
	 		\label{fig:ProjPlan}
  		%	\Description{After the main deployment a phase of TD repayment follows.}
		\end{figure}

		\textit{Processing:} The uniqueness of the TAP framework is that the repayment of these intentionally incurred TD items is part of the project plan. 
		No project can be finished before the TD tickets that have been incurred during this project are processed. 
		The project plan does not end after reaching the deadline and keeping to the tight schedule. 
		Instead, the development team and architects prioritize the TD tickets of this project in a prioritization meeting. %Streichkandidat (notfalls)
		The prioritization is discussed and decided by a majority vote and does not follow a strict procedure.
		In a subsequent project phase, the TD repayment phase, the developers process the project's TD tickets alternately with the project's remaining requirement tickets, as shown in \Cref{fig:ProjPlan}. 
		Thus, the responsibility for the TD repayment is transferred to project management because to finish their project is in the project manager's interest.
		In this way, the TAP framework integrates the management of intentional TD into the project management process. 
		
		\textit{Goal:} 
		First, evaluating different solution options in advance raises the overall awareness only to incur TD intentionally.
        Consequently, unintentionally incurred TD is prevented. 
		Second, making the project managers responsible for the TD accumulated during their project decreases their willingness to incur TD.
		In other words, this creates a feedback loop, and both the team and the project manager change their behavior towards TD incurrence.
		Third, project managers and IT managers get an overview of accumulating TD of a particular project while still running.
		The IT managers can intervene or adjust the project plan and the following projects early on. 
		Finally, by this approach, intentionally incurred TD is repaid timely.

	    \textit{Example:} 
		For the example from \Cref{section:introduction}, this may mean that hard-coded VAT rates are adjusted in the code rather than refactored to a flexible solution to meet the schedule. 
		As a result of this decision, the developers record the corresponding TD ticket to introduce a centralized tax variable. This TD will be repaid in an orderly manner in the ``TD repayment phase'' of the ``VAT adjustment'' project and after the law comes into force. 
        In contrast, discussion of the various solution options may show that the low effort makes centralization the better option, and TD are prevented from the beginning.

		\subsubsection  {Deconstruction}
		\label{section:deconstructionTickets}
		A ticket of the deconstruction category is a special TD ticket that cannot be processed during the project. 
		After implementing a new solution, the legacy code sometimes needs to be kept parallel for the time being.
		These tickets then comprise the deconstruction of this legacy code when it is no longer needed. 
		They can also comprise the deletion of deprecated code parts after all consumers of these parts are detached. 
		Therefore, this category only comprises intentional TD that cannot be repaid immediately, i.e., during the TD repayment phase of its project.
	
		\textit{Recording:} These tickets must contain additional information about when the deconstruction can take place. 
		
		\textit{Processing:} The deconstruction ticket must be repaid as soon as possible as part of the contingent for functional requirements.
		
		\textit{Goal:} The primary purpose of these tickets is to avoid code cluttering and improve code comprehension. 
		
	    \textit{Example:} 
		An example is a need for a new business rule for the following business year. 
		The application should not be deployed on New Year's Eve to minimize deployment risks. 
		Hence, the application will be deployed in advance and must contain a date-driven switch to choose between the old and new business rules. 
		When the old business rule is no longer needed, it can be deconstructed. 
		%Streichen?
		%The ``point in time'' in this ticket would be January, 1 of the respective next year.

		\subsection  {Scope}
		\label{section:FramworkScope}
		The initial scope of this framework is the IT unit that developed the framework.
		Nevertheless, other units and companies may adopt the framework or at least relevant parts of it. 
		Relevant parts are, in particular, the distinction in processing intentional TD as TD tickets and unintentional TD as part of maintenance tickets, the immediate recording of intentional TD, and the processing of intentional TD as part of the project that  incurred them. 
		
		We think this framework will particularly help units that are subject to tremendous time pressure.
		The framework was used in an organizational setting of agile project management, including agile processes and project management elements, including project deadlines. 
		Regardless, we expect the framework to help in all organizational environments where a timeline is relevant, i.e., project or release deadlines.
		
		The framework may not be suitable for agile product development without timeline constraints.
		Additionally, the problem caused by accumulated TD must be large enough to make the initial effort to organize the backlog worthwhile.
		
	\section{CASE STUDY DESIGN} 
	% Was ist der theoretische Rahmen? Das Framework oder das related Work?
	\label{section:csmethod}
	The goal of this research was the presentation and evaluation of the TAP framework.
	The framework's presentation took place in \Cref{section:framework}.  
	The evaluation was based on the case study design, according to Runeson \cite{Runeson2012}.
	
	\subsection{Research Questions}
	\label{section:rq}
	As described in \Cref{section:introduction}, the two main problems concerning TD are preventing TD and dealing with TD despite having a tight schedule and fixed deadlines. 
	The TAP framework provides a possible solution found and developed in practice for these problems. 
	With subsequent research questions (RQ), this paper provided an evaluation of the TAP framework. 
	We evaluated especially the framework's feasibility and effectiveness regarding the problems mentioned above.\\
	
    \textit{(\textbf{RQ 1}) Framework Application}
    \begin{itemize}
    \setlength{\parskip}{1pt}
        \item [] \textit{(\textbf{RQ 1.1}) Do practitioners find the TAP framework reasonable?} 
        \item [] \textit{(\textbf{RQ 1.2}) Are the processes of the TAP framework feasible in practice?}
    \end{itemize}
	This is the basis for all other evaluations as it is not reasonable to analyze the framework's impact further should it not be feasible in practice. 
	We used a team survey of the observed IT unit's members and ticket statistics to evaluate these questions.\\
	
    \textit{(\textbf{RQ 2}) TAP Framework's perceived Effects}
    \begin{itemize}
    \setlength{\parskip}{1pt}
        \item [] \textit{(\textbf{RQ 2.1}) Is the use of the TAP framework associated with a raised awareness for the incurrence of TD?}
        \item [] \textit{(\textbf{RQ 2.2}) Is the use of the TAP framework associated with a more conscious incurrence of TD items?}
        \item [] \textit{(\textbf{RQ 2.3}) Is the use of the TAP framework associated with a better overview of TD items?}
        \item [] \textit{(\textbf{RQ 2.4}) Is the use of the TAP framework associated with a timelier repayment of TD items?}
    \end{itemize}
	These aspects are the TAP framework's goals, as presented in \Cref{section:framework}. 
	The evaluation of these questions shall show whether these goals were reached. 
	To evaluate these questions, we used ticket statistics and surveyed the members of the observed and a comparison unit. \\
	
    \textit{(\textbf{RQ 3}) TAP Framework's perceived Benefits and Drawbacks}
    \begin{itemize}
    \setlength{\parskip}{1pt}
        \item [] \textit{(\textbf{RQ 3.1}) Do practitioners observe that TD can be prevented by the adoption of the TAP framework?} 
        \item [] \textit{(\textbf{RQ 3.2}) Are there other benefits or drawbacks arising from the adoption of the TAP framework?}
        \item [] \textit{(\textbf{RQ 3.3}) Do practitioners find these benefits justify the additional effort?}
    \end{itemize}
	The purpose of these questions is to gather the TAP framework's impact.
	The main goal of TD prevention is evaluated, as well as secondary benefits concerning team discussion, decision-making process, and generating an overview of TD and maintenance tasks. 
	To answer these RQs, we used descriptive statistics, and correlations between the survey variables for effectiveness and benefits to support the descriptive statistics.
	Furthermore, we asked for the justification of the additional effort in relation to the benefits to substantiate the framework's feasibility.\\

    \textit{(\textbf{RQ 4}) Management Perspective}
    \begin{itemize}
    \setlength{\parskip}{1pt}
        \item [] \textit{(\textbf{RQ 4.1}) Do the managers have an overview of their IT systems' TD?} 
      %  \item [] \textit{(\textbf{RQ4.2}) What are the effects of the overview of TD from a management perspective?}
        \item [] \textit{(\textbf{RQ 4.2}) What are the benefits of the TD overview from a management perspective?}
        %\item [] \textit{(\textbf{RQ4.3}) What are the resulting benefits of this overview?}
    \end{itemize}
    The success of the TAP framework depends largely on the management's support as the additional effort must be approved by them.
    Therefore, we evaluated the effect the framework and especially the generated TD overview has on the management.
    
    \subsection{Case and Units of Analysis}
	\label{section:CaseDescription}
    % Definitions
	Following the definitions of Runeson \cite{Runeson2012}, we conducted a comparative case study with two units of analysis - the observed IT unit that developed and adopted the TAP framework and the comparison unit that did not.
	The composition of the observed unit and the comparison unit are presented in \Cref{tab:unitcomp}.
	The comparison unit is slightly smaller but is led by the same unit manager and follows a similar organizational structure.
	
	The focus of this case study was the evaluation of the framework's effects on the observed IT unit.
	The case study protocol, as well as all data collection and evaluation details and the data itself can be accessed in the additional material\footnote{ \url{https://doi.org/10.5281/zenodo.5788222}}.

    %Context
    \textbf{\textit{Context.}}
        The IT units are part of a German publishing house with more than 9000 employees worldwide.
        This is an established company publishing content and advertisements in magazines, websites, and mobile apps.
        Both IT units follow a Scrum approach and develop and maintain software for internal use. 
        They are both dealing with a substantial amount of legacy code.
	
    \textbf{\textit{Unit of analysis.}} 
    % Unit of analysis
        The observed unit develops and evolves systems responsible for the marketing of advertisements. 
        The main competitors are Google and Facebook.
        This leads to often changing systems due to new advertising formats and the merging of companies to get a bigger market share.
        Additionally, the changes are usually time-sensitive to be first in the market or adhere to the business year when merging companies.
    
        The unit develops about 5-10 medium-large systems and the interfaces between these systems.
        The development tasks are organized in agile-managed projects. 
        The unit comprises three organizational teams with two cross-cutting Scrum teams. 
        The three organizational teams comprise:
        \begin{itemize}
        \setlength{\parskip}{1pt}
            \item Business Analysis: The members are responsible for the requirements engineering and project management.
            \item Development: The members focus solely on the systems' development.
            \item Operations and Support: The members are responsible for customer communication, quality assurance, and basic server support.  
        \end{itemize}
        Each Scrum team is responsible for specific tasks depending on the required technology's expertise.
        The business analysts are acting as product owners in the Scrum processes.

        The unit of analysis was not chosen but given, as this IT unit developed the framework for itself.
        
    \textbf{\textit{Comparison Unit of Analysis.}} 
    % Comparative Unit of analysis
        The comparison unit develops and evolves systems responsible for internal company tasks, like support for human resources, janitor services, company intranet, or canteen.
        
        The unit develops about 20 small- to medium-size systems and their interfaces.
        Like the observed unit, the comparison unit uses agile projects for their development and has to reach project deadlines. 
        On the one hand, in some of their projects, the time pressure is less stringent than in the observed unit.
        On the other hand, some of their projects have to adhere to legal requirements, including hard-set deadlines.
        
        Furthermore, the IT management particularly encourages this unit to explore new technologies, leading to additional risk of TD and legacy code.
        The comparison unit consists of exactly one team, which is both the organizational team and the Scrum team.
        This team has one business analyst who is also the architect and the product owner in the Scrum process.
        
        The comparison unit of analysis was chosen due to their availability as the manager of both units supports this study. 
        Furthermore, both units follow a similar organizational approach, i.e., agile-managed projects.
     
	\subsection{Data Collection}
	\label{section:dataCollection}
	To answer RQ 1.2 and RQ 2.4, we evaluated the recording and processing statistics for TD and maintenance tickets.  
	RQs 1 to 3 were answered by the results of the team survey. 
	Finally, a follow-up survey investigated the TAP framework's effects on the IT managers, which answered RQs 4.
	Both surveys targeted participants from the observed unit and the comparison unit.

	\subsubsection{Ticket Statistics}
	\label{section:DCTicketStatistics}
	We collected data from all project backlogs of the observed unit and identified the tickets under consideration by tags and epic affiliation.
	All tickets in the backlog were labeled according to their category at the time of data collection. 
	524 maintenance tickets and 141 TD tickets were identified.
	
	In consultation with the teams' architects, we validated the raw data and removed old or invalid tickets. 
	Furthermore, we narrowed down the tickets to the period from January 2018 to March 2020.
	236 maintenance tickets and 102 TD tickets remained for the evaluation.
	
	We did not evaluate the statistics for maintenance projects and deconstruction tickets due to insufficient data, e.g., only ten deconstruction tickets exist between 2018 and 2021.

	\subsubsection{Team Survey}
	\label{section:SurveyMethod}
	We chose the method of a survey to benefit from the input of all team members willing to share their experiences. 
	For RQ's 1 to 3, we asked corresponding questions in this questionnaire.

	%\subsubsection{Survey Participants} %paragaph
	%\label{section:SurveyParticipants}
	\textbf{\textit{Survey Participants.}} 
    %Survey Participants
    	We asked all members of the observed and comparison IT unit to fill out the questionnaire. 
    	%The comparison unit is led by the same unit manager which increases the comparability.
    	The comparison unit's results allowed us to validate the descriptive statistics.
    	The participants were asked which team they belong to, but due to works council regulations, they were not asked which role (e.g., architect, manager) they inhibit.
    	The response rate to the survey is shown in \Cref{tab:unitcomp}. 

    	\begin{table}
    	    \centering
    	    \footnotesize
    		\begin{tabular}{lcccc} % p{2,1cm}
    			\toprule
    				&\multicolumn{2}{c}{observed unit} 	&\multicolumn{2}{c}{comparison unit}\\ 
    							&members&particip. 	&members&particip.\\ 
    			\midrule 
    			manager 			& 2& n/a 			& 2 	& n/a \\ 
    			architect		 	& 2 & n/a 			& 1$^{\mathrm{a}}$    & n/a \\ 
    			business analyst	& 8 & 6   			& 1$^{\mathrm{a}}$ 	& 1 \\
     			developer 			& 15& 9   			& 5 	& 4 \\ 
    			operations  		& 5 & 2   			& 0 	& 0\\ 
    			\midrule 
    			sum 	& 32& 17  & 8 	& 5\\ 
    			response rate	& & 53\%  &  	& 62\%\\ 
     			\bottomrule
     			\multicolumn{5}{l}{$^{\mathrm{a}}$architect and business analyst are the same person}
    		\end{tabular}
    		\caption{Composition of units and team survey participants}
    		\label{tab:unitcomp}
    	\end{table}
	%\subsubsection{Questionnaire Construction}  %paragaph
	%\label{section:QuestionnaireCnstruction}
    \textbf{\textit{Questionnaire Construction.}}
    %Questionnaire Construction 
        To avoid misunderstandings, the questionnaire started with short information about the term TD. 
    	For statistical reasons, background information was queried, e.g., the participant's team membership.
    	The questionnaire was divided into two main parts: (I) the framework's assessment and (II) the framework's effects and benefits. 
    	
    	Part (I) showed the framework's practical feasibility and provided answers to RQ 1.1 and RQ 1.2. 
    	For every ticket category, the same set of questions regarding the reasonableness and feasibility was asked. 
    	Each part started with short information about the ticket category.
    	Part (I) was not filled out by the comparison unit as these participants did not use the framework. 
    
    	Part (II) comprised five subsections with questions regarding: 
    	\begin{itemize}
    	\setlength{\parskip}{1pt}
    	    \item TD awareness (RQ 2.1)
    	    \item comparison of optimal and sub-optimal solutions (RQ 2.2)
    	    \item overview of TD (RQ 2.3)
    	    \item observed benefits of the comparison and overview (RQ 2.4, RQ 3.1 and RQ 3.2)
    	    \item justification of the additional effort (RQ 3.3)
    	\end{itemize}
    	
    	All parts included a set of assertions the participants were asked to validate using the following Likert scale\cite{Joshi2015}: applies - rather applies -  rather does not apply - does not apply - cannot answer. 
    	The \textit{cannot answer} option was given as some questions could not be answered by all members of the unit.
        The participants were explicitly asked only to use this option in these cases. 
    	
    	Finally, two more open questions at the end allowed the participants to point out good parts of the TAP framework and parts needing improvement.
    
    	Two developers familiar with the TAP framework filled out a first questionnaire as a pilot test. 
    	Based on this, we reduced and focused the questions. 
    	One IT manager gave additional feedback on the optimized questionnaire's construction, which we incorporated.
       
	\subsubsection{Follow-up Management Survey}
	\label{section:ManagementSurveyMethod}
	 During a first presentation of the evaluation results, the management mentioned that the framework provides a valuable overview of the TD and maintenance tasks.
	 This perception contradicts what most of the survey participants assumed of their management (see \Cref{fig:WhoHasOverview}).
	 Since the advantages of TD management for IT managers are not often researched, we investigated this in more detail.
	 
    \textit{\textbf{Survey Participants.}} 
        The survey participants were only the three involved line managers: the unit manager of both units, the team manager of two teams of the observed unit, and the team manager of the comparison unit. 
        All managers have personnel responsibilities for the teams involved and have a good and broad overview of the business goals.
        Due to the small sample size, we could not generalize the results.
        However, the evaluation provided qualitative input exploitable in further research.
    
    \textit{\textbf{Questionnaire Construction.}} 
        We asked the managers four open questions via email to gather qualitative data.
        The questions were related to
        \begin{itemize}
        \setlength{\parskip}{1pt}
            \item the overview they have of their system's TD, 
            \item the effects this has 
            \begin{itemize}
            \setlength{\parskip}{1pt}
                \item on their project pipeline, 
                \item on TD repayment and prevention, and 
                \item on the communication with their customers. 
            \end{itemize}
        \end{itemize}

	\subsection{Data Analysis}  %                   ?????????????????                  
	\label{section:dataAnalysis}
	
    \subsubsection{Ticket Statistics}
	\label{section:DATicketStatistics}
	To evaluate RQ 1.2 and RQ 2.4, we presented the number of tickets in terms of categories and time as descriptive statistics for the period from January 2018 to March 2020.

	\subsubsection{Team Survey}  %paragaph
	\label{section:SurveyEvaluation}
	For the questionnaire's evaluation, we used the statistics software SPSS\footnote{\url{https://www.ibm.com/dede/analytics/spss-statistics-software}} for closed questions and the qualitative research software MaxQDA\footnote{\url{https://www.maxqda.de/}} for open questions.
	
	First, we created descriptive statistics of all closed questions for an exploratory data analysis.
	For simplification and analysis purposes, we dichotomized the answers for the questions of part (II). 
	This means we summarized the answers \textit{applies} and \textit{rather applies} as well as \textit{rather does not apply} and \textit{does not apply}. 
	We present all values as a percentage to align the output of the observed unit and the smaller comparison unit. 
	For the sake of clarity, we only show the \textit{applies} answers in the following figures. 
	
	Second, we developed hypotheses for RQ 2 based on this exploration and evaluated the significance of the hypotheses. 
    We used the Mann-Whitney U-Test \cite{Mann1947} because it can be applied to not normally distributed data on an ordinal scale.
	We present only significant findings.
	
	Third, we analyzed correlations between survey participants who incur TD consciously and the benefits they assess.
	%The comparison unit could not assess the benefits (RQ 3.2 and RQ 3.3.) because they mostly do not compare different solution options.
	%Their answers must be interpreted as expected benefits, and we could not evaluate the differences between the units regarding the benefits. 
	%Instead, we evaluated these correlations regardless of their team affiliation. 
	We used the $\phi$-coefficient of Pearson's $\tilde{\chi}^2$-correlation \cite{Cohen1988,Pearson1900} for dichotomous variables and assumed a significant correlation from a significance level of 0.05, i.e., 5\%.
	The effect size is interpreted as medium if it is higher than at 0.3 and as strong if higher than 0.6.
	
	Last, two researchers used open coding to evaluate the open questions.
	
	\subsubsection{Follow-up Management Survey }  %paragaph
	\label{section:FollowupSurveyEvaluation}
	Two researchers evaluated the management questionnaire's results using open coding with MaxQDA. 
	Due to the limited number of participants, we did not count codes.
	Instead, we provided some qualitative insights of the line managers of the observed and comparison unit.
	  
	\section{RESULTS}
	\label{section:results}

		\subsection  {Ticket Statistics}
		\label{subsection:TicketStat}

\begin{figure}
		 		\centering
	 			\begin{tabular}{@{}c@{}}
					\subfigure[Maintenance tickets by time]
					{	\label{fig:MTbyTime} 
						\includegraphics[width=0.45\textwidth]{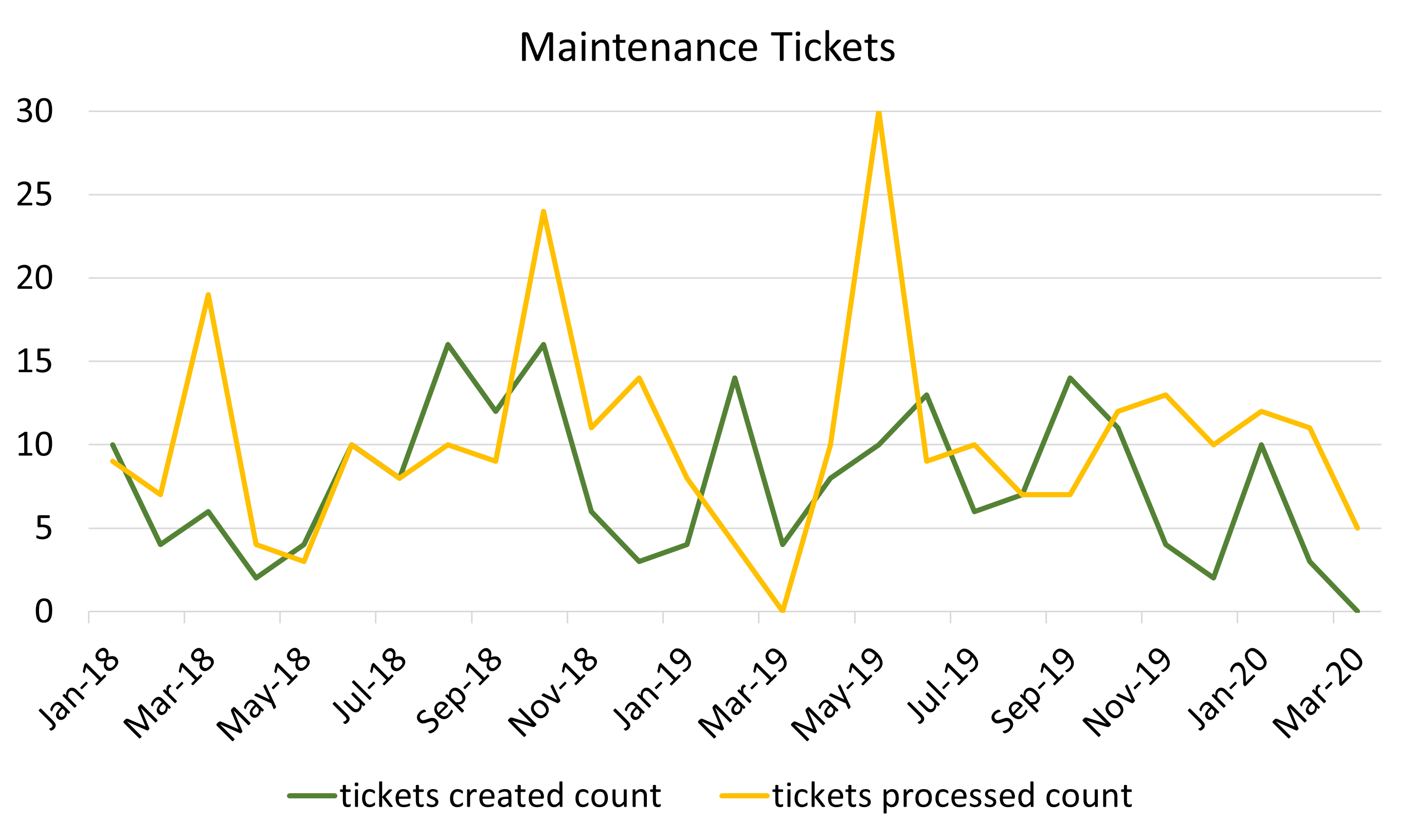}
					}\\
					\subfigure[TD tickets by time]
					{	\label{fig:TDTbyTime}
						\includegraphics[width=0.45\textwidth]{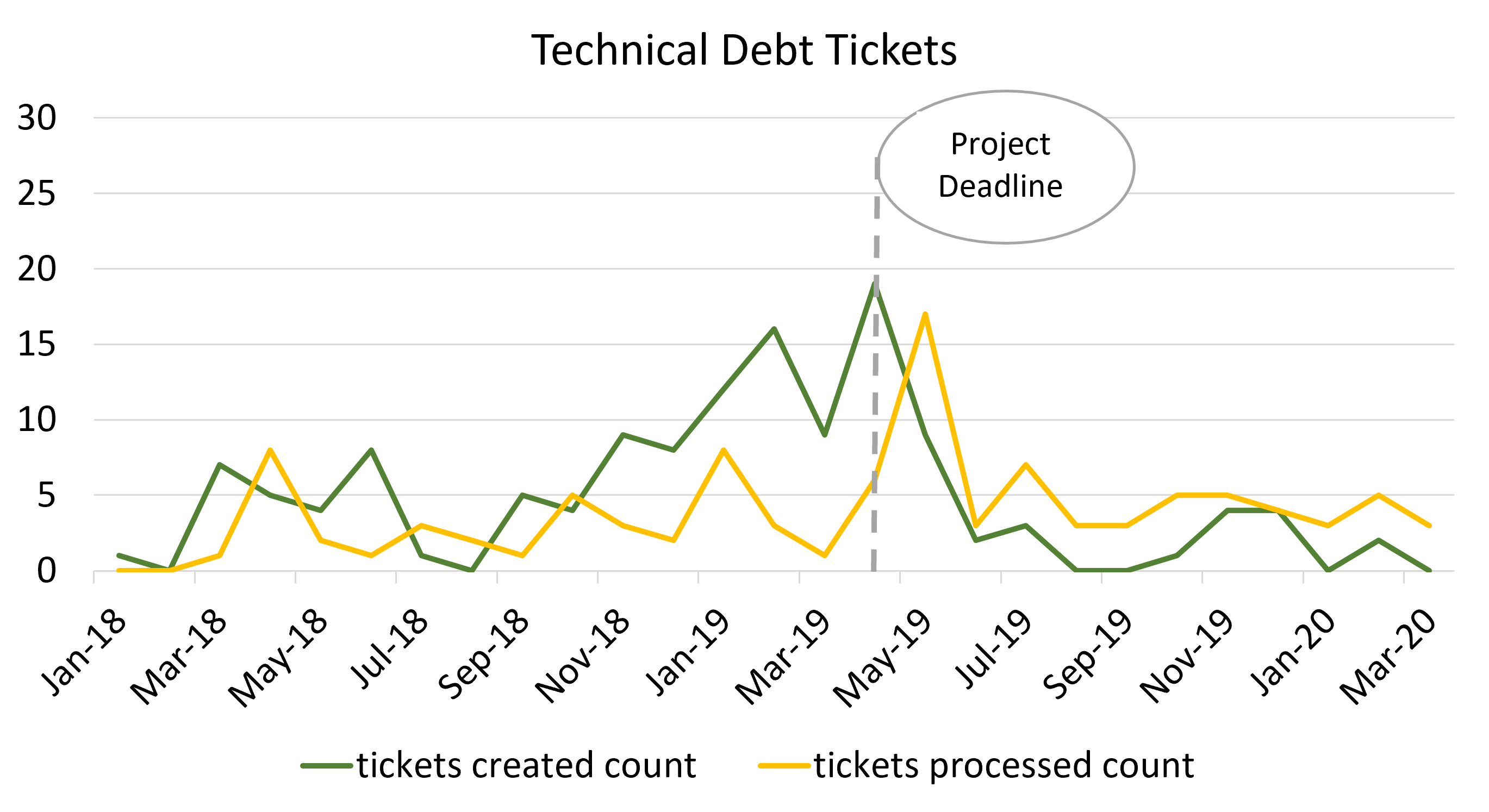}
					}
	 			\end{tabular}
				\caption{Ticket statistics with a project deadline at the end of April 2019}
				\label{fig:TicketbyTime}
	  			%\Description{Maintenance tickets are constantly created and done. TD tickets are accumulated in spring and repaid in summer and autumn.}
			\end{figure}

		The timelines in \Cref{fig:MTbyTime,fig:TDTbyTime} show the development of ticket counts over time to answer RQ 1.2. (feasibility) and RQ 2.4 (timely repayment). 
		The maintenance tickets were created and processed continually.
		The timeline for TD tickets shows a peak in incurred TD in spring 2019 due to a project with a tight deadline. 
		No maintenance or TD tickets were processed in March 2019. 
		Accordingly, the TD tickets as well as the maintenance tickets show a processing peak after the reached deadline at the end of April 2019. 
		Initially, only the TD tickets with the highest priority were processed.
		From July to November 2019, the TD tickets with a lower priority were processed, meaning that this project was still not completed (see \Cref{fig:ProjPlan}). 
		Other projects that started in parallel caused this long duration because development capacities had to be divided between the projects.

		\subsection  {Team Survey}
		\label{subsection:survey}
 %---------------------------------------------------------------------------------
			\subsubsection  {Assessment of the TAP Framework}
		    \label{subsection:ResultsAssesment} 
			
\begin{figure*} 
			    \centering
	 			\begin{tabular}{@{}ccc@{}}
					\subfigure[Is the recording of the ticket reasonable in general?]
					{	\label{fig:RecordingRes} 
						\includegraphics[width=0.3\textwidth]{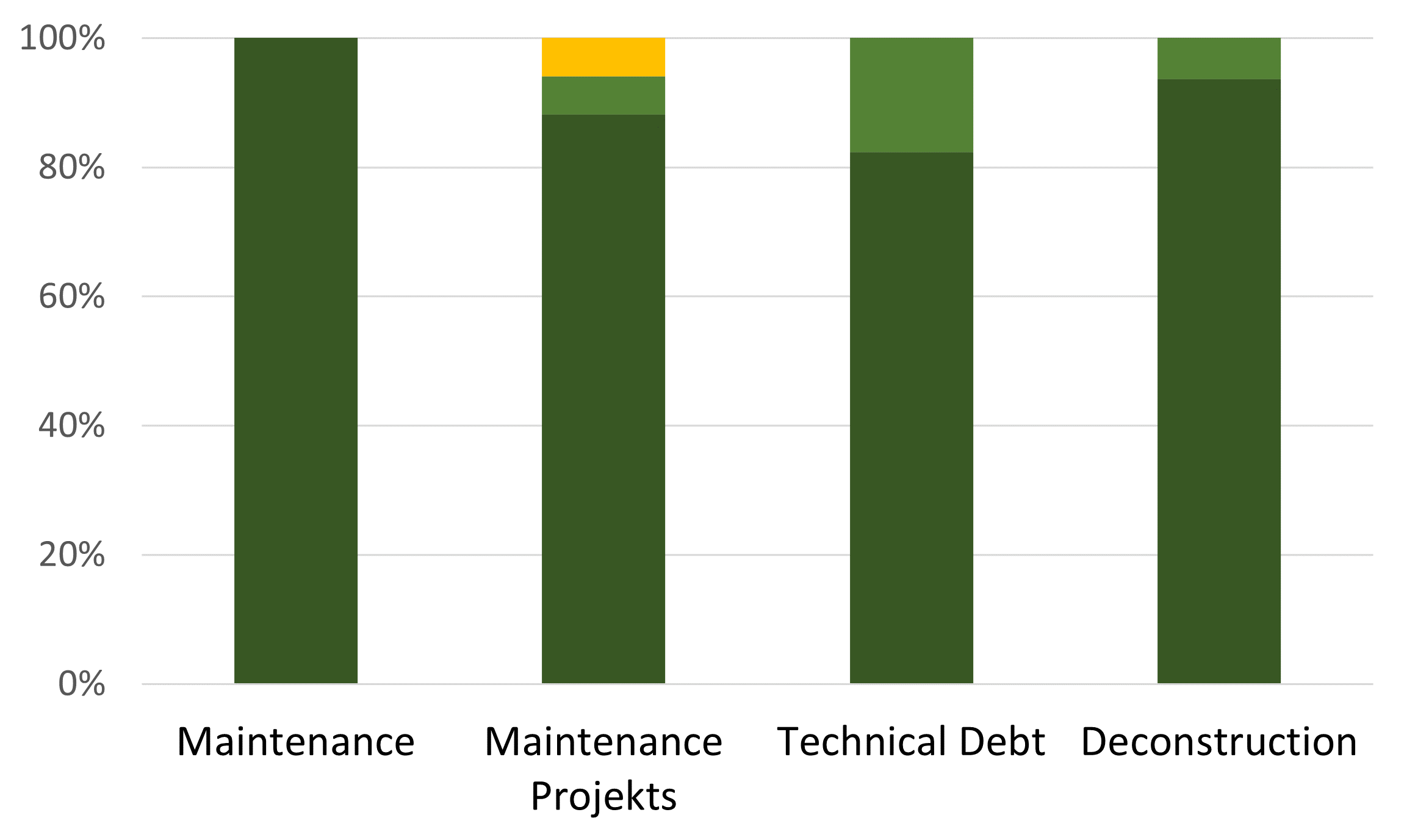}
					}&
					\subfigure[Is the procedure for recording the ticket reasonable?]
					{	\label{fig:RecordingProcRes} 
						\includegraphics[width=0.3\textwidth]{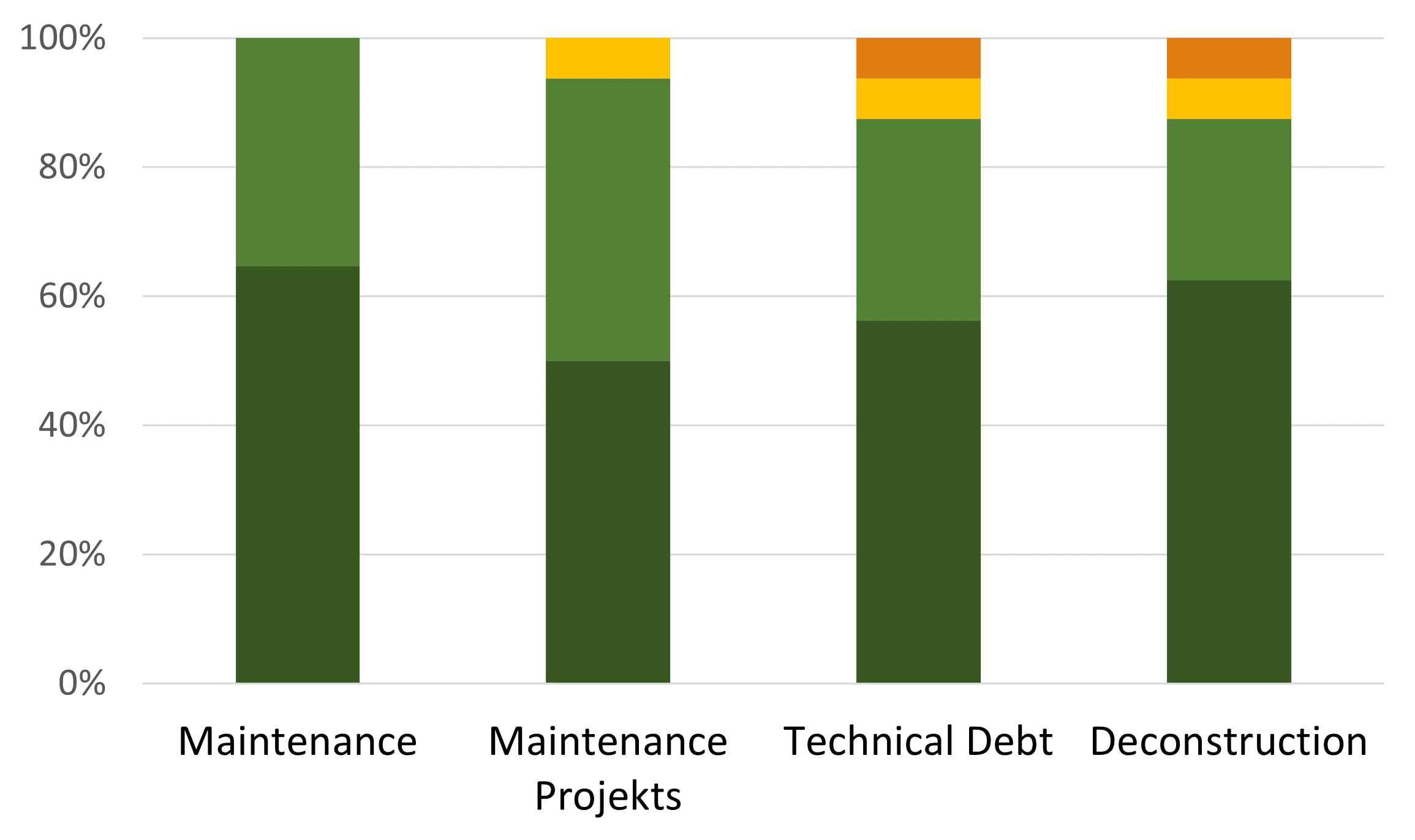}
					}&
					\subfigure[Does the procedure for recording the tickets work?]
					{	\label{fig:RecordingProcWork} 
						\includegraphics[width=0.3\textwidth]{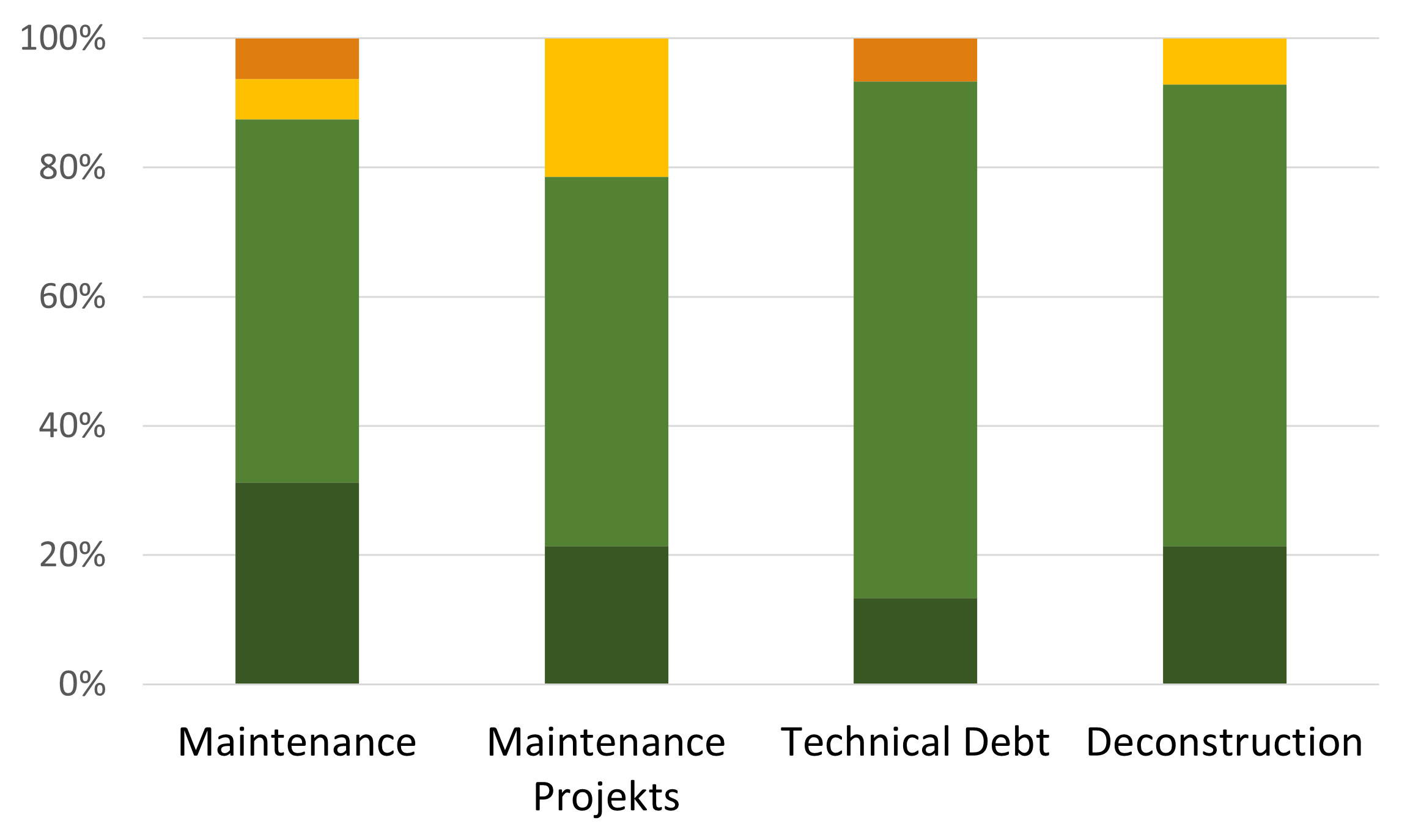}
					} \\
					\subfigure[Is the processing of the ticket reasonable in general?]
					{	\label{fig:ProcessingRes} 
						\includegraphics[width=0.3\textwidth]{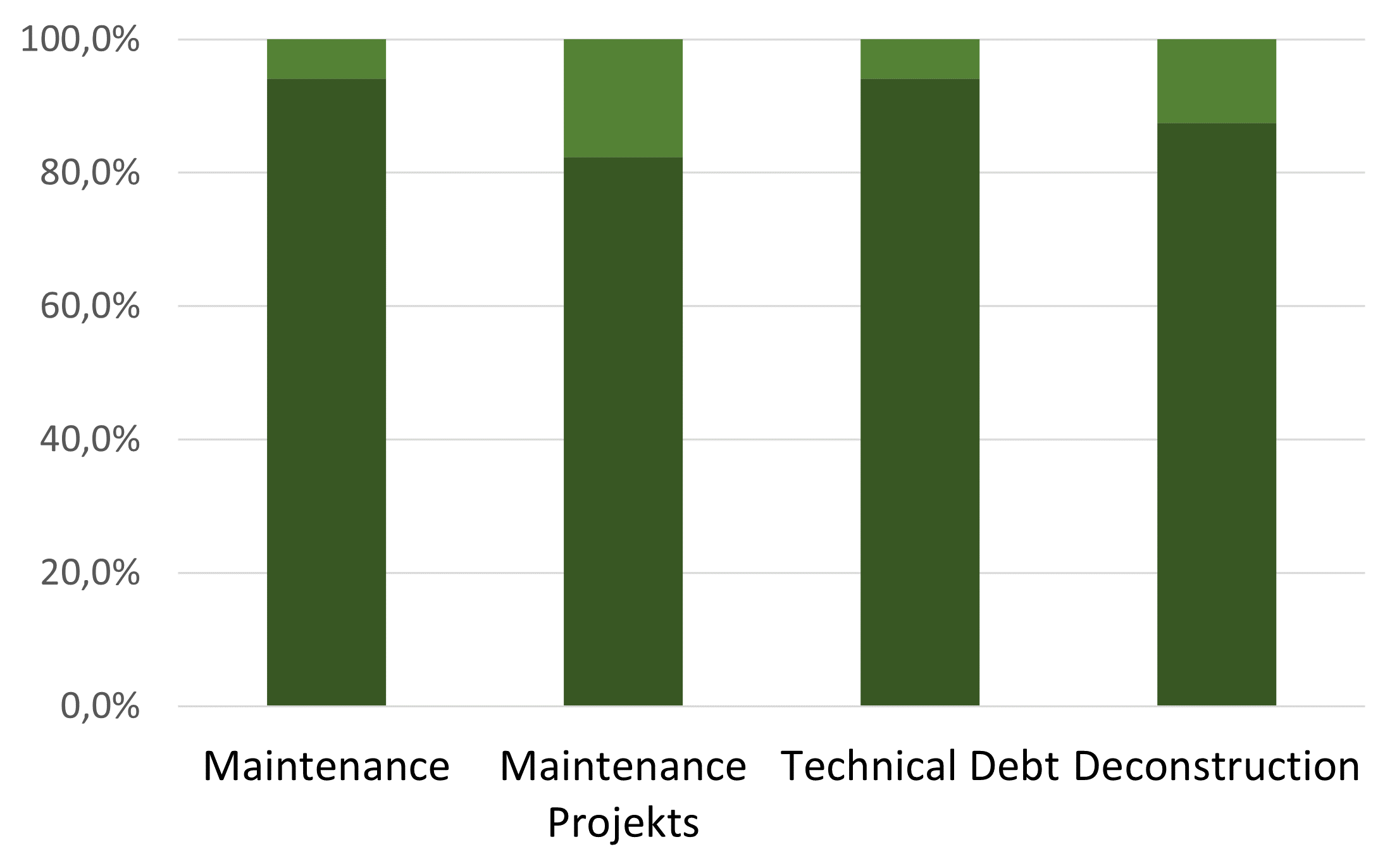}
					}&
					\subfigure[Is the procedure for processing the ticket reasonable?]
					{	\label{fig:ProcessingProcRes} 
						\includegraphics[width=0.3\textwidth]{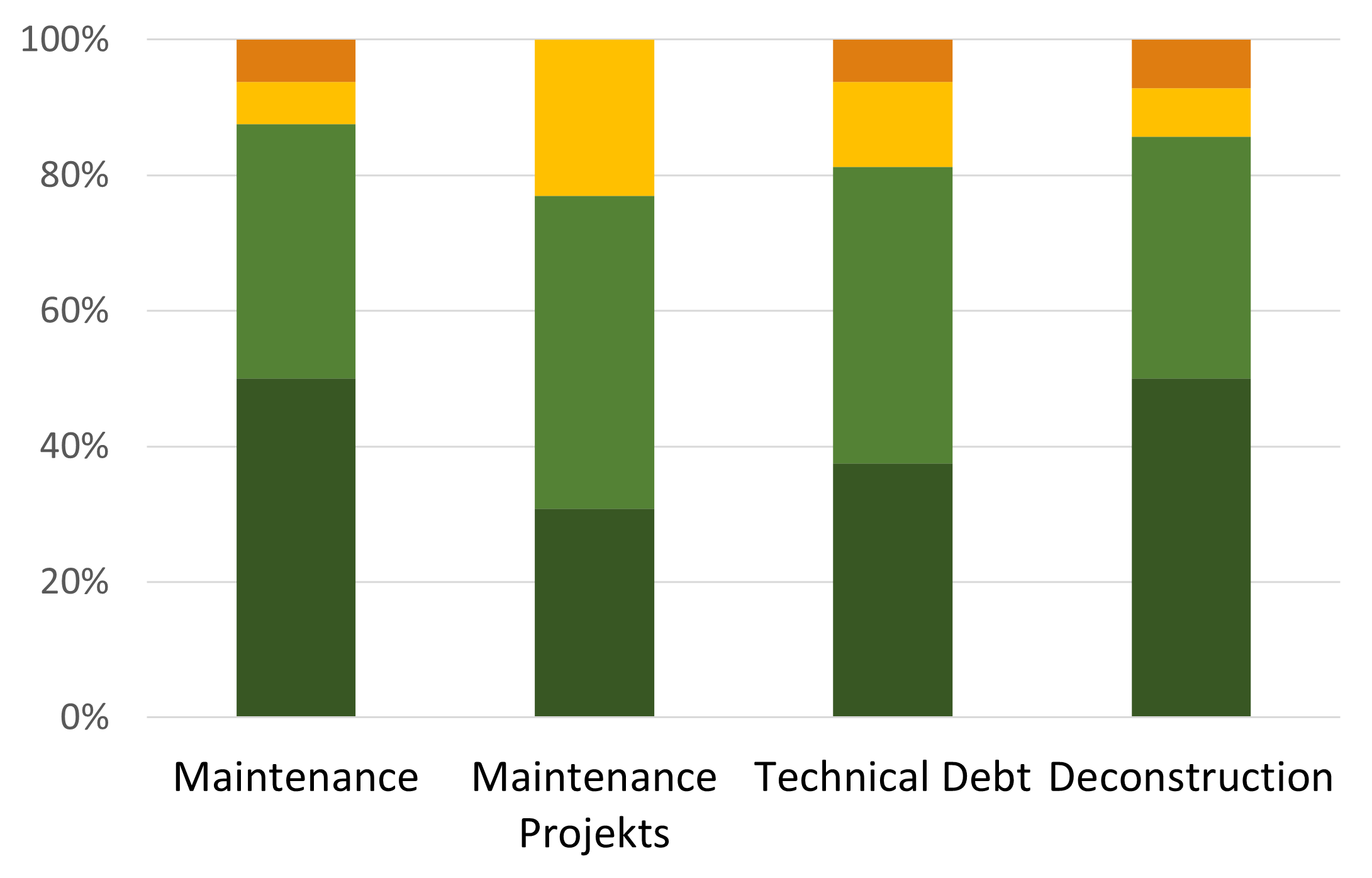}
					}&
					\subfigure[Does the procedure for processing the tickets work?]
					{	\label{fig:ProcessingProcWork} 
						\includegraphics[width=0.3\textwidth]{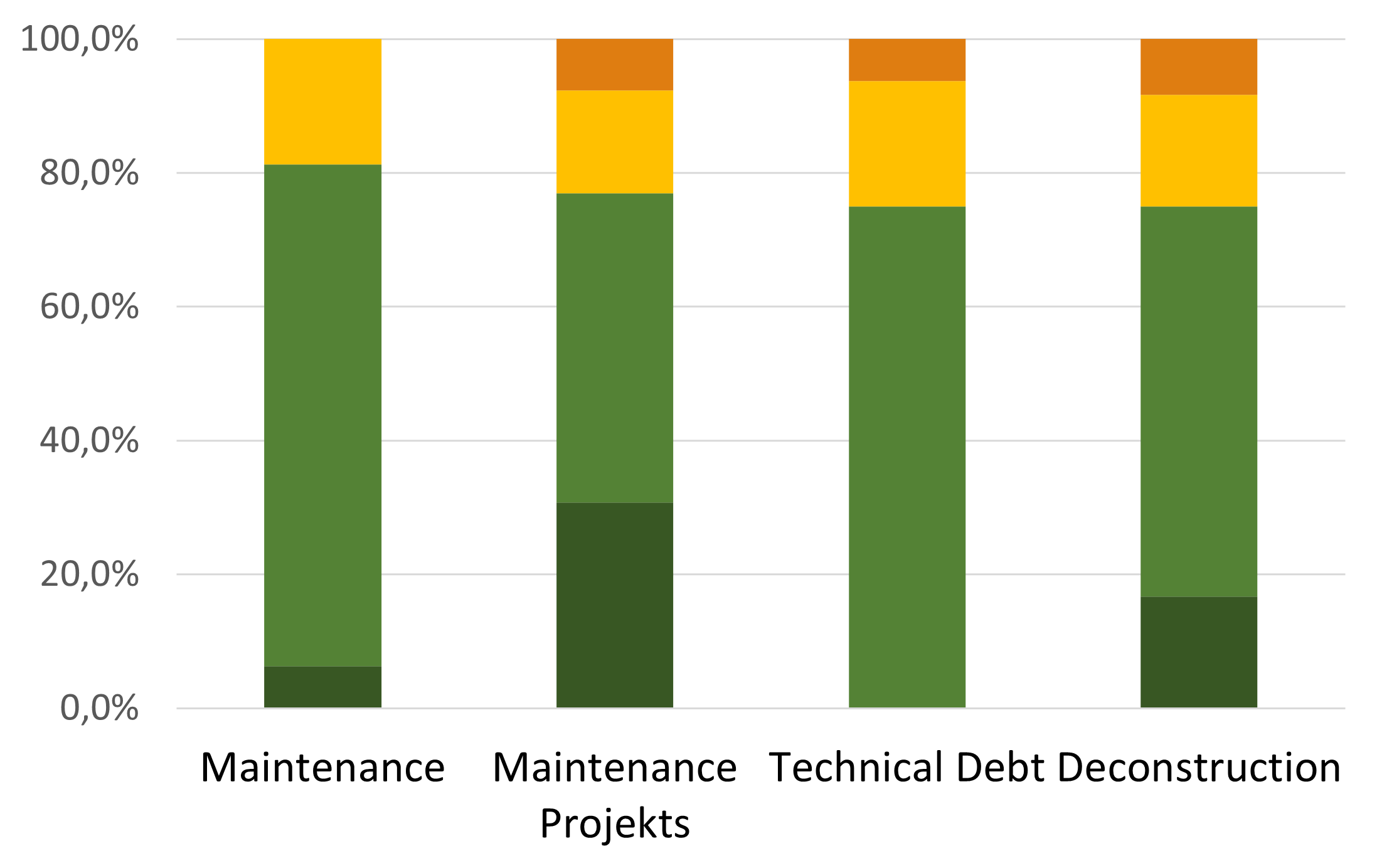}
					}\\
 				
					\multicolumn{3}{c}{\includegraphics[width=0.4\textwidth]{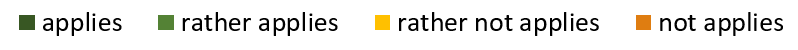} } \\
	
	 			\end{tabular}

			    \caption{Appropriateness of the TAP Framework - Descriptive Statistics}
			\end{figure*}

			Most survey participants agreed that recording and processing all four types of tickets are generally reasonable  (\Cref{fig:RecordingRes,fig:ProcessingRes}). 
			Most of them also agreed that the framework's procedures for recording and processing are reasonable (\Cref{fig:RecordingProcRes,fig:ProcessingProcRes}). 
			When asked whether the framework worked as intended, the agreement decreased, but most of the survey participants still stated that the procedures work well and are helpful (\Cref{fig:RecordingProcWork,fig:ProcessingProcWork}).
			For RQ 1.1 and RQ 1.2, these results mean the TAP framework is reasonable and feasible in practice.

			\begin{figure*}
			    \begin{tabular}{@{}cc@{}}
    				% \subfigure[Do the survey participants recognize that TD is taken on while incurring them?]
    				% 	{	\label{fig:RecTD} 
    				% 		\includegraphics[width=0.48\textwidth]{"TDRecognition"}
    			 % 			%\Description{No difference between the two units}
    				% 	}&
    				\subfigure[To what extent are the sub-optimal and the optimal solution compared?]
    					{	\label{fig:CompareQvO}
    						\includegraphics[width=0.48\linewidth]{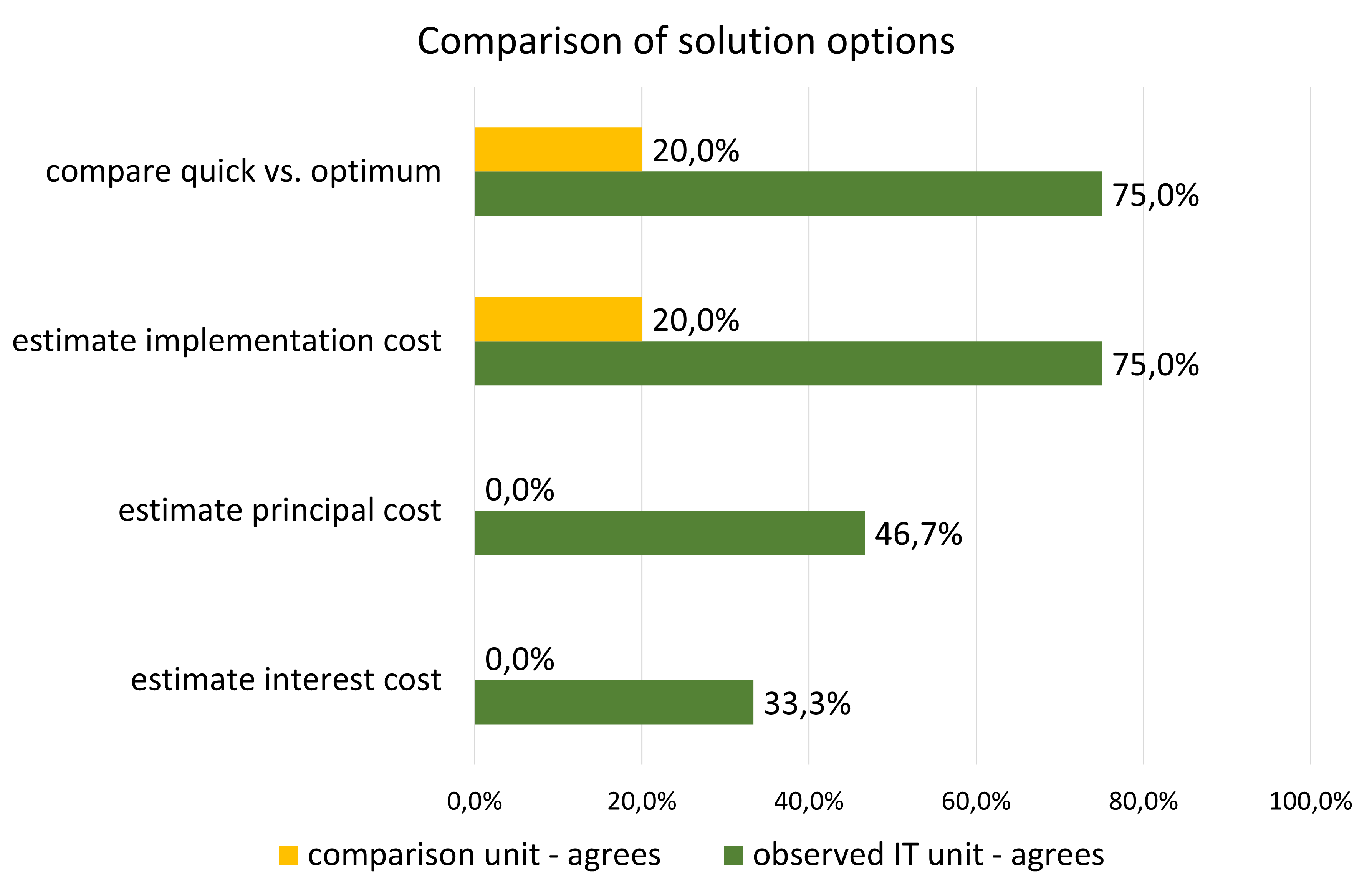}
    			  			%\Description{A significant difference can be seen.}
    					}&
    			    \subfigure[Which stakeholder has an overview of all TD items?]
    					{	\label{fig:WhoHasOverview}
    						\includegraphics[width=0.48\textwidth]{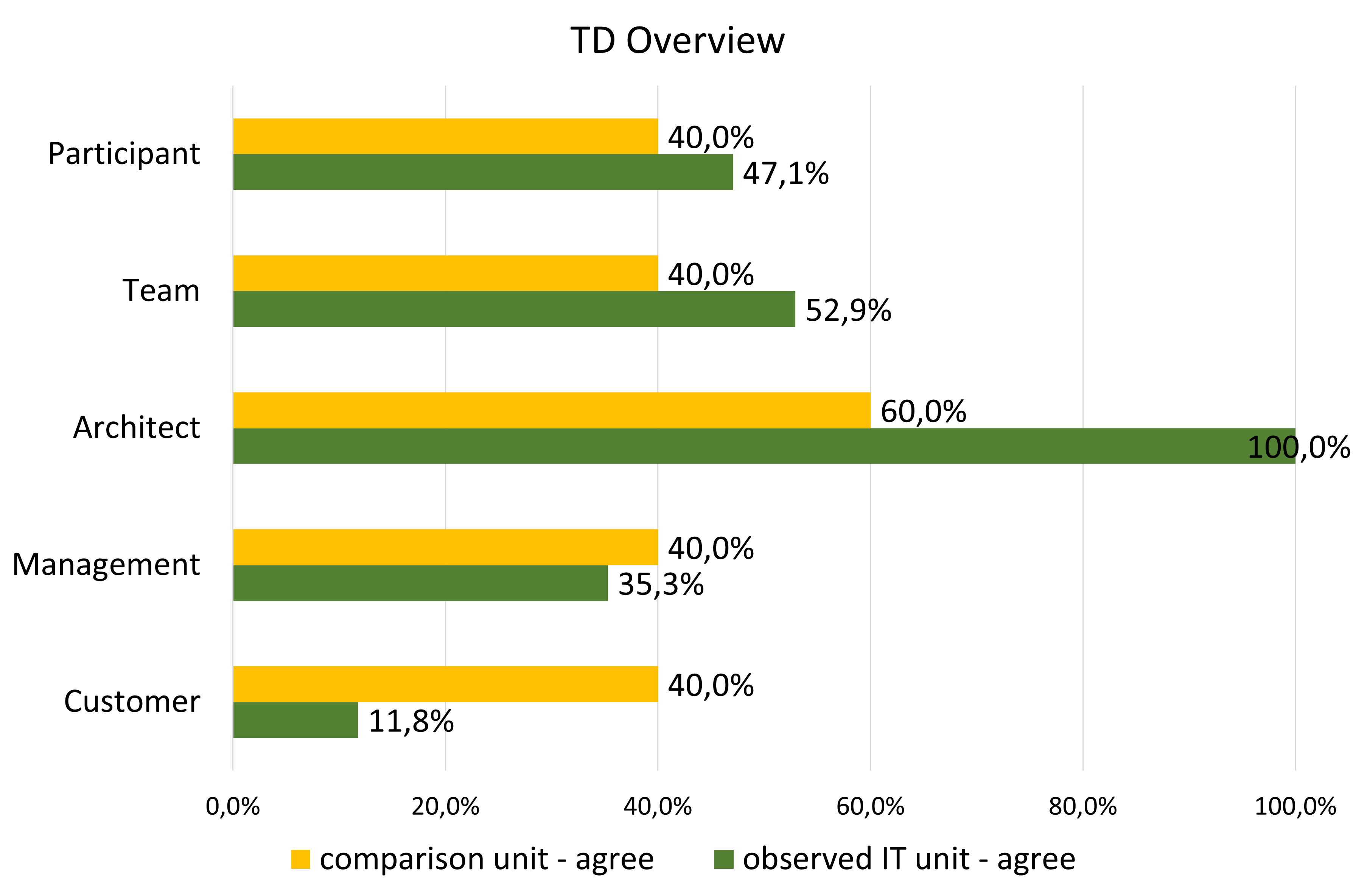}
    			  			%\Description{A difference between groups can be seen, but that difference is not significant.}
    					}\\
    				\subfigure[What benefits of the comparison can be found?]
    					{	\label{fig:BenefitsDecision}
    						\includegraphics[width=0.48\textwidth]{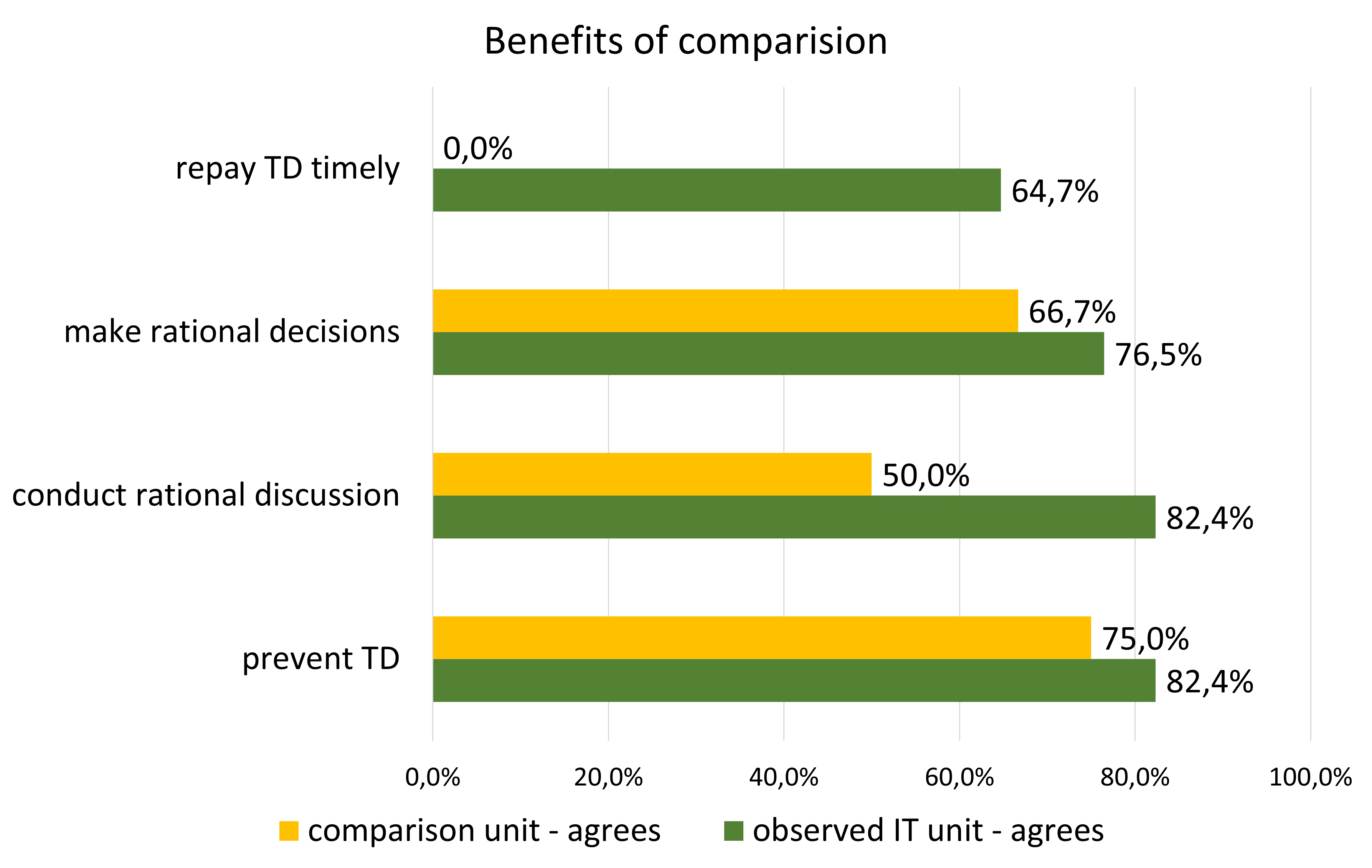}
    			  			%\Description{A difference between groups can be seen, but that difference is not significant.}
    				
    					}&
    				\subfigure[What benefits of the overview can be found?]
    					{	\label{fig:BenefitsOverview}
    						\includegraphics[width=0.48\textwidth]{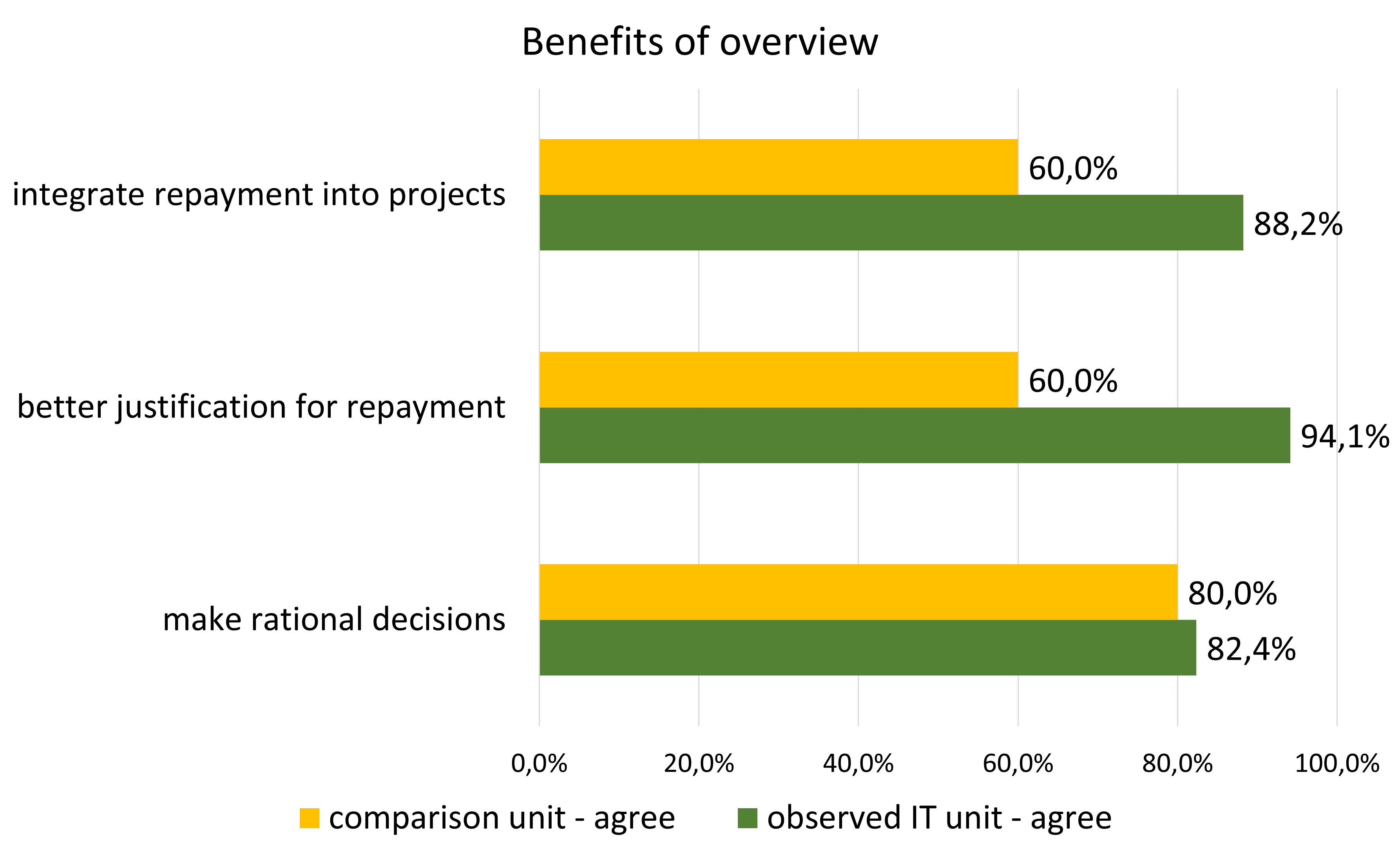}
    			  			%\Description{A difference between groups can be seen, but that difference is not significant.}
    					}
    % 				\subfigure[Do the survey participants think the extra effort for this is justified?]
    % 					{	\label{fig:EffortJustified}
    % 						\includegraphics[width=0.48\textwidth]{"EffortJustified"}
    % 			  			%\Description{All units find the extra effort to be justified.}
    % 					}
	 		    \end{tabular}
			    \caption{Usability, Effects, and Benefits of the TAP Framework - Descriptive Statistics}
			\end{figure*}
			
			\begin{figure} % [ht]
         		\centering
        		\includegraphics[width=0.48\textwidth]{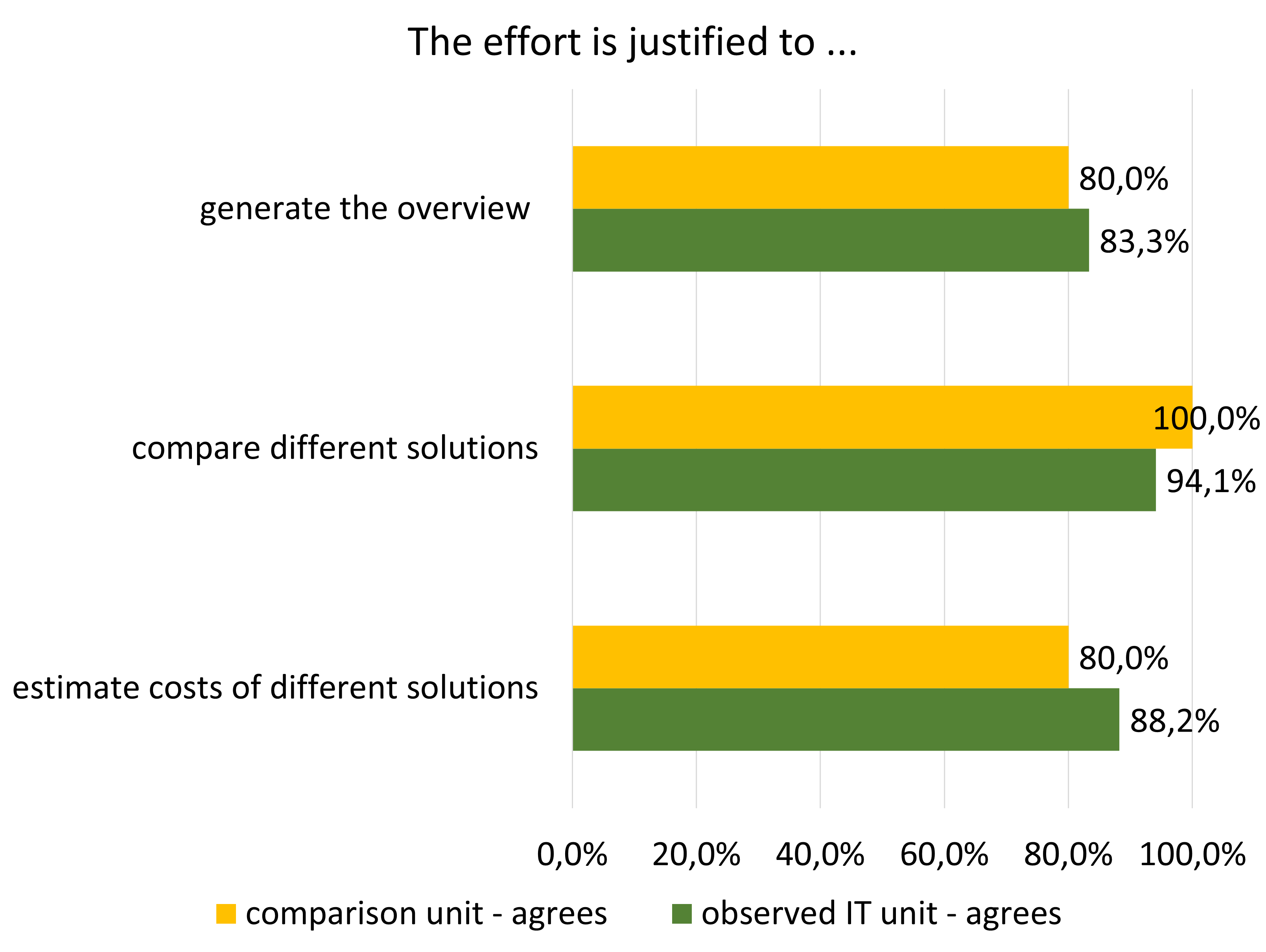}
        		\caption{Do the survey participants think the extra effort for this is justified?}
         		\label{fig:EffortJustified}
          	%	\Description{After the main deployment a phase of TD repayment follows.}
        	\end{figure}
% 			\begin{figure*} 
% 			\centering
% 			\begin{tabular}{@{}cc@{}}
% 				\subfigure[Do the survey participants think the extra-effort for this is justified?]
% 					{	\label{fig:EffortJustified}
% 						\includegraphics[width=0.48\textwidth]{"EffortJustified"}
% 			  			%\Description{All units find the extra effort to be justified.}
% 					}
% 	 		\end{tabular}

% 			\caption{Effects and benefits of the Framework - Descriptive Statistics}
% 			\end{figure*}

        \subsubsection  {Perceived Effects of the TAP Framework}  
		\label{subsection:ResultsEffects}

            First, we asked survey participants whether they and their team typically recognize that they incurred TD items before or after their incurrence.
			The goal of this question was to identify whether the participants were aware of the incurrence to answer RQ 2.1.
			%As can be seen in \Cref{fig:RecTD} the units show no significant difference. 
			With more than 75.0\% agreement, both units stated that they usually recognized it when TD items were incurred.
			The results are not significantly different. 

			Second, the participants were asked whether they compare different solution options and estimate respective implementation costs, principal, and interest rates for incurring TD. 
			To answer RQ 2.2, we assumed a conscious and intentional decision to incur TD if the units compared different solution options.
			Regarding this question, an apparent difference between the units could be seen in \Cref{fig:CompareQvO}. 
			75.0\% of the observed unit's participants agreed to compare sub-optimal and optimal (or more) solutions and to calculate implementation cost for this comparison. 
			Only 20.0\% of the participants agreed to compare options and estimate implementation costs for the comparison unit. 
			Furthermore, many participants of the observed unit agreed that they estimate principal (46.7\%) and interest costs (33.3\%), while this was not the case for the comparison unit.
			
			These observations led to the following research hypothesis (H\textsubscript{R}) in comparison to the respective null hypothesis (H\textsubscript{0}):
			
    		\textit{\textbf{H\textsubscript{R.1}:}} The observed unit compares sub-optimal and optimal solutions more frequently than the comparison unit. 
    		(H\textsubscript{0.1}: The observed unit does not show different behavior than the comparison unit in terms of comparing sub-optimal and optimal solutions.)
    		
    		\textit{\textbf{H\textsubscript{R.2-4}:}} The observed unit estimates the implementation (and principal and interest) costs for different solutions more frequently than the comparison unit. 
    		(H\textsubscript{0.2-4}: The observed unit does not show different behavior than the comparison unit in terms of estimating the implementation (and principal and interest) costs for different solutions.)
			
			\begin{table} 
			    \centering
        		\begin{tabular}{llc}
        			\toprule
        			H\textsubscript{R} 	& variable & exact significance\\  
        			\midrule 
        			H\textsubscript{R.1} & comparison		    & 0.058 \\ 
        			H\textsubscript{R.2} & implementation cost  & 0.048 \\ 
        			H\textsubscript{R.3} & principal cost	    & 0.002 \\
         			H\textsubscript{R.4} & interest cost 		& 0.015 \\ 
         			\bottomrule
        		\end{tabular}
        		\caption{Significance of H\textsubscript{R}-hypothesis}
        		\label{tab:MannWhitney}
        	\end{table}
			
			\Cref{tab:MannWhitney} shows the significance values of these hypotheses measured with the Mann-Whitney U-Test. 
			The exact significance for H\textsubscript{R.2-4} hypotheses is below 0.05. 
			Consequently, these hypotheses can be accepted as they are significant to the 5\% significance level. 
			This means the possibility these hypotheses are wrong is lower than 5\%.
			The exact significance for H\textsubscript{R.1} hypothesis is 0.058 and a little higher than 5\%.
		    H\textsubscript{R.1} cannot be accepted.
		    While these statistics give an additional indication on the difference between the units, we have to keep the low number of data points in mind.
		    
			In summary, this means the framework's adoption is associated with a more conscious comparison of different solutions options. 
			In particular, the estimate of the respective costs of the solution options differs significantly between these two units.
			
            Third, we asked whether the framework's adoption generated an overview of TD items to answer RQ 2.3.
            To evaluate this, we asked the participants which stakeholder has an overview of all TD. 
            \Cref{fig:WhoHasOverview} shows that in the observed IT unit, the architects had an overview of all TD (100\% agreement). 
            The customer was not expected to know the system's TD (11.8\% agreement).
            In the case of the comparison unit, it was unclear which stakeholder had an overview of TD.
 % ----------------------------------------------------------------       
% Correlation - Tabelle muss hier stehen, damit sie auf der richtigen Seite angezeigt wird - warum auch immer ...        
			\begin{table*} [t] 
			    \centering
			    %\footnotesize
				
				\begin{tabular}{l|ll|ll|ll}%p{1,5 cm}
						\toprule
							& \multicolumn{2}{c|}{TD prevention}  
							& \multicolumn{2}{c|}{rational discussions}  
							& \multicolumn{2}{c}{rational decisions}  
							\\
							& $\phi$-coeff & signif. 
							& $\phi$-coeff & signif. 
							& $\phi$-coeff & signif. 
							\\ 
						\midrule 
						comparing solutions & 
						0.419 &  0.061 & 0.303 & 0.176  & \textbf{0.623}  & \textbf{0.007}\\
						estimating implementation costs &
						0.357 &  0.110 &  \textbf{0.471} &  \textbf{0.035} &  \textbf{0.535} & \textbf{0.020}\\
						estimating principal costs &
						0.394 & 0.086 & \textbf{0.456} & \textbf{0.047} & \textbf{0.495} & \textbf{0.036}\\  
						estimating interests costs &
						0.309 & 0.179 & 0.357 & 0.120 & 0.385 & 0.103\\ 
    			 	    \bottomrule
				\end{tabular}%sub
				\caption{Correlations of comparison and estimations vs. benefits \\
				(Significant correlations are highlighted)}
				\label{tab:Correlations}
			\end{table*}
 % ----------------------------------------------------------------  
			
	\subsubsection  {Perceived Benefits of the TAP Framework}  
		\label{subsection:FrameworkBenefits}

			To answer RQ 3.1 and 3.2, we asked the participants what benefits they expected (comparison unit) or assessed (observed unit) of the comparison of different solution options and, thus, the conscious incurrence of TD.
			\Cref{fig:BenefitsDecision} shows that both units mostly agreed that this comparison could prevent TD. 
			Especially the observed unit assessed that discussions were led more rational (82.4\% agreement) and more rational decisions were made (76.5\% agreement). 
			Timely repayment (RQ 2.3) of TD was not expected by the comparison unit but assessed by many participants of the observed unit (64.7\% agreement). 
			This difference may partly be due to a misunderstanding that the observed unit was unconsciously assessing the framework's benefits and not just the comparison's benefits.
			The units show no relevant differences in benefits, which means no hypotheses can be derived.  
			
            To answer RQ 3.2, the participants were asked what benefits they expected or observed from the generated overview of TD items.
            The observed unit's participants noted that integrating the TD items into existing or upcoming projects is easier with an overview (88.2\%).
            Furthermore, the overview gives a better justification for the repayment (94.1\%) (see \Cref{fig:BenefitsOverview}). 
            Only 60\% of the comparison unit expected the overview's benefits, which is a noticeable smaller but still not significant amount.
            Both units expected or observed more rational decisions due to the generation of the overview.

			Finally, to answer RQ 3.3, the participants were asked whether the benefits justified the effort for comparing and estimating the options and generating the TD overview.
			Both units mostly thought the effort was justified (\textgreater80.0\% agreement). 
			Again, the units show no apparent differences (see \Cref{fig:EffortJustified}). 
		
 %---------------------------------------------------------------------------------
			
			\subsubsection  {Correlations between effects and benefits}% of the TAP framework} 
	    	\label{subsection:ResultsCorrelation}
	    	
			As shown in the previous section, the main difference caused by the framework was that the observed team compared different solution options and estimated the costs for new TD tickets.
			We observed different benefits that may result from this comparison and estimations: more rational discussions, more rational decisions, and the prevention of TD.
			To support the observation and explore the possibility of a causal effect between the comparison and estimations and these benefits, we calculated the correlations between them.
			We also calculated the significance of these correlations and assumed a correlation if the significance was less than 0.05. 
			
			%As pointed out in \Cref{section:SurveyEvaluation}, the framework's benefits could not be assessed by the comparison unit.
			%Therefore, to identify the comparison's benefits, we analyzed correlations between all survey participants that compare different solutions and the benefits they assessed. 

		 	%\Cref{tab:Correlations} shows correlations that can be found. Significant correlations are highlighted.

			\Cref{tab:Correlations} shows significant correlations of medium strength between the perceived benefit of making more rational decisions and the observed effects of comparing solutions, estimating implementation costs, and estimating principal costs. 
			The benefit of leading more rational discussions has significant correlations of medium strength to the estimation of implementation and principal costs. 
			
			Following our observations and the correlations, we can associate the comparison and estimations lead with more rational decisions. 
			Furthermore, we associate the estimation of implementation and principal costs with more rational discussions.
			Both associations seem to be valid assumptions since they are also logically derivable.
			
			The correlations do not support associations for TD prevention or the estimation of interest costs.
			%The missing correlations for interest costs may be due to the fact that only one third of the participant estimate these costs.
			%The missing correlations for TD prevention leads to the assumption that that the perceived prevention of TD is not  directly caused by the comparison and estimations but may be caused by other factors.
			
			As we did not have a significant difference between the units related to the overview of TD, we did not calculate correlations for the overview's perceived benefits.

			\subsubsection  {Open Questions}  
	    	\label{subsection:OpenQuestions}
			Most comments were only made one time and are single opinions. 
			Two points were mentioned three times each. 
			
			First, ``TD is a topic'' expresses that it was already helpful to talk about the topic of TD and not to ignore it.
			The code is related to the goal of raising TD awareness.
			One participant pointed out: 
			``\textit{I actually see the main advantage of conscious handling of the topic in the matter as such, because otherwise, it is a topic that is almost ignored in everyday development.}'' 
			
			Second, ``Sometimes TD is OK'' shows that there were situations in which participants considered retaining TD reasonable, e.g., in legacy systems. 
			Another participant stated: 
			``\textit{If a system is to be replaced, then I consider technical debts to be justifiable if they disappear after 1-2 years anyway}''. 
			
			A third point was mentioned two times: ``Evaluation is difficult and subjective''. 
			This topic refers to the research topic of TD prioritization. 
			The participants mentioned that it is sometimes hard to decide which TD should be repaid first and whether a particular item is TD at all, especially if the architectural rules seem to be open for interpretation. 
			One participant phrased it as follows: 
			``\textit{Even if the ``teaching book'' says that the current implementation does not correspond to the architectural specifications, it still seems to be a personal opinion that decides on the fundamental correctness of the specification.}''  
            The other participant mentioned that TD is easy to define as a deviation from the norm or rules. However, it is ``\textit{very difficult to evaluate this deviation and work out the importance of fixing this debt.}''
				%In erster Linie finde ich, dass man technische Schulde objektiv sehr gut definieren kann (etwas fällt aus einer Vorgabe / weicht von der Norm ab) andererseits ist es sehr schwer, diese Abweichung zu bewerten und die Wichtigkeit der Behebung dieser Schulden zu erarbeiten. (Was ist besonders erfolgreich, Pos. 3)
			
			One participant also mentioned that the ``framework raises the understanding for each other'': ``\textit{I'm thinking of greater domain understanding in development and greater technical understanding for business analysts.}''
				%Ich denke da an ein größeres fachliches Verständnis in der Entwicklung und ein größeres technisches Verständnis in der Beratung. Denn obwohl wir immer wieder sagen, dass wir ein Team sind, bleiben wir doch häufig getrennt und gerade die unterschiedlichen Tickets tragen dazu bei, dass die Unterschiede der Teamteile sichtbar werden.  (Was ist besonders erfolgreich, Pos. 4)
			Regarding the ``more rational discussions'', one participant states: ``\textit{If the process is consciously lived and ingrained in people's minds, it reduces energy-sapping discussions about fixing technical debt.}''	
				%Wird der Prozess bewusst gelebt und in den Köpfen verankert, so reduziert es kräftezehrende Diskussionen um die Behebung technischer Schulden. ( Vor-Nachteile bewusster Umgang, Pos. 8)
			Another participant summarizes these effects of the comparison: ``\textit{The conscious handling of different possible solutions promotes cooperation, mutual understanding, and knowledge exchange in the team and thus generally contributes to the development of stable and accepted software. }''
			Lastly, one participant acknowledges the ``importance of TD prevention'': ``\textit{The incurrence of debt should be reduced, not the process for working it off optimized.}''
				%Die Aufnahme von Schulden sollte reduziert werden, nicht das Verfahren zur Abarbeitung optimiert.  (Was sollte optimiert werden, Pos. 4)
				%mittelfristig könnte es dazu führen, dass in den Projekten weniger "TS" entstehen, da die Abarbeitung, wenn man an dieser Stelle ordentlich vorgeht - auch Zeit/Kosten verursacht - und das könnte man sich sparen, wenn man "vorne am FlieÃŸband diese Zeit einsetzt um mehr Qualität liefern zu können ( Vor-Nachteile bewusster Umgang, Pos. 9)
				  
			The following \textbf{drawbacks} were mentioned by the participants. 
			One participant stated that the categories' distinction and the separation of their responsibilities might lead to a separation of the team. 
			``\textit{Although we keep saying that we are one team, we often stay separate and the different tickets in particular help make the differences between the team parts visible}''.
			Another participant thought the framework ``\textit{(slightly) contradicts the agile principle of providing a functioning solution with company value as quickly as possible. 
			At the same time, however, it levels the stable ground so that we can continue to work on new requirements quickly and without production errors. 
			The advantage outweighs}''.
			One participant of the operations team finds that in their team TD is not a topic and that ``\textit{This is a concept that takes place in the development cosmos.}'' 
			He recommends that the operations team should be involved in the repayment of TD. 
			 %Bei AQM ist das Thema "technische Schulden" überhaupt nicht präsent. Das ist ein Begriff der sich im Entwicklungskosmos abspielt. (Was sollte optimiert werden, Pos. 3)
			 %Es sollte präsenter sein, AQM sollte in den Aufbau von techn, Schulden involviert werden.  (Was sollte optimiert werden, Pos. 3)

		\subsection  {Follow-up Management Survey}
		\label{subsection:managementSurvey}
		In this survey, we asked the IT managers of the observed and comparison unit for their perspective on their units' TD management. 
		The unit manager responsible for both units answered for each unit separately.
		Thus, we got four different responses from three participants: 
		\begin{itemize}
        \setlength{\parskip}{1pt}
		    \item team manager of the observed unit (TMO), participant A
		    \item team manager of the comparison unit (TMC), participant B
		    \item unit manager of the observed unit (UMO), participant C
		    \item unit manager of the comparison unit (UMC), participant C
		\end{itemize}
		
        \textit{\textbf{Overview of TD.}} 
        The TMO and UMO both stated they had an overview of their TD. 
        Both managers relied on a weekly meeting with the architects to establish this overview. 
        In these meetings, architects and IT managers reviewed the architecture, planed the evolution of the system and architecture, and discussed the TD and maintenance projects. 
        The architects prepared the meeting by providing a prioritized pipeline of maintenance projects and an architectural evaluation of the TD items.
        The TMC and UMC did not have a specific mechanism to create an overview of their TD. Still, they remarked that they had a ``\textit{holistic project pipeline that identifies technical debts as such and made it possible to plan corresponding topics in an integrative manner.}'' 
        The unit manager, who experienced the strategies of both units, remarked regarding the comparison unit: ``\textit{In terms of future operational responsibility, I would feel more comfortable if the technical debt in the project was managed continuously and transparently.}'' 
        He stated his concern that ``\textit{TD does not become clear until the handover of operations or even during ongoing operations.}''
        
        \textit{\textbf{Project Pipeline.}} 
        For the TMO and UMO, the over\-view had a direct impact on the planning of maintenance projects. 
        It also helped to plan capacities and estimate project duration realistically. 
        The UMO even mentioned the possibility that ``new projects start later.''
        The TMC stated that TD ``\textit{projects are given a further measurable criterion for their urgency.}'' 
        The TMC and UMC further mentioned that it was still a problem to prioritize ``between business requirements and technical necessity'', while the TMO or UMO did not mention this problem.
        
        \textit{\textbf{TD Repayment Activities.}}
        The TMO and UMO planed their TD repayment activities during their weekly meeting with the architects. 
        During these meetings, it became apparent in which cases it was sensible just to accept the TD, e.g., when the system will be re-engineered soon. 
        The UMO further mentioned the effect the visibility may have had for the teams: 
        ``\textit{The existing debts become visible, and in the end, developers and product owners have to decide whether further debts are justifiable or realistically repayable in the foreseeable future.}'' 
        The TMC focused on TD awareness and implied that the overview led to enhanced awareness.
        He stated that ``\textit{TD awareness leads to TD avoidance}'' and that TD awareness ``\textit{supports the case for projects and measures to reduce this debt}''. 
        
        \textit{\textbf{Communication with Customers.}}
        The TMO and UMO said the pipeline has helped make effects visible, making it easier to communicate with customers.  
        Project durations were better understood and accepted. 
        The UMO mentioned that ``\textit{it also reflects the customers' interest in a) being able to use the right features on time, but b) also to use a stable and future-proof system in the medium to the long term}''. 
        He remarked that this also included communication with other stakeholders, e.g., higher management or users. 
        The TMC saw the same benefits but pointed out that the understanding also depended on the customer's technical know-how.
        
	\section{DISCUSSION}
	\label{section:discussion}
	In this section, we will finally answer our RQs from \Cref{section:rq} regarding the TAP framework developed by practitioners.
	We used different methods to answer our research questions.
	\Cref{tab:ResultsOverview} gives an overview of which method and figure will be used to answer the respective questions.
		
		\begin{table*}[h]
			\centering
			\begin{tabular}[h]{p{7,5cm}lp{1.2cm}l} %p{8 cm}p{2 cm}p{1,5 cm}p{4 cm}}
			\toprule
			RQ & Method &  Results Section & Figure/Table \\ 
			\toprule
			
			 RQ 1.1 (TAP  framework's reasonability) & team survey 
			 & \ref{subsection:ResultsAssesment} %\nameref{subsection:ResultsAssesment} 
			 & \cref{fig:ProcessingRes,fig:ProcessingProcRes,fig:RecordingRes,fig:RecordingProcRes}   
			 \\ 
			 RQ 1.2 (TAP  framework's feasibility) & team survey
			    & \ref{subsection:ResultsAssesment} %\nameref{subsection:ResultsAssesment}  
			    & \cref{fig:ProcessingProcWork,fig:RecordingProcWork}   \\ 
			 & tickets statistics 
			    & \ref{subsection:TicketStat} %\nameref{subsection:TicketStat}
			    & \cref{fig:TicketbyTime}   \\ 
			 \midrule
			 
			 RQ 2.1 (raised awareness) & team survey 
			    & \ref{subsection:ResultsEffects} %\nameref{subsection:ResultsEffects}   
			    &     \\
    	      &  
    	        & \ref{subsection:OpenQuestions} %\nameref{subsection:OpenQuestions}  
    	        &     \\  
			  & management survey 
			    &  \ref{subsection:managementSurvey} %\nameref{subsection:managementSurvey}
			    &     \\  
			 
			 RQ 2.2 (conscious incurrence) & team survey  
			    & \ref{subsection:ResultsEffects} %\nameref{subsection:ResultsEffects}  
			    & \cref{fig:CompareQvO},{\crefname{table}{tab.}{tab.} \cref{tab:MannWhitney}}   \\ 
			 &  
			    & \ref{subsection:OpenQuestions} %\nameref{subsection:OpenQuestions}  
			    &   \\ 
			 RQ 2.3 (TD items' overview) & team survey  
			    & \ref{subsection:ResultsEffects} %\nameref{subsection:ResultsEffects}   
			    & \cref{fig:WhoHasOverview}  \\ 
			 %& management survey 
			 %   & \ref{subsection:managementSurvey} %\nameref{subsection:managementSurvey} 
			 %   &   \\ 
			 RQ 2.4 (timely repayment) & team survey 
			    & \ref{subsection:FrameworkBenefits} %\nameref{subsection:FrameworkBenefits} 
			    & \cref{fig:BenefitsDecision}  \\ 
			 & 
			    & \ref{subsection:OpenQuestions} %\nameref{subsection:OpenQuestions}
			    &   \\ 
			 & tickets statistics  
			    & \ref{subsection:TicketStat} %\nameref{subsection:TicketStat}  
			    & \cref{fig:TicketbyTime}  \\ 
			 
			 \midrule
			 RQ 3.1 (TD prevention) & team survey 
			    &  \ref{subsection:FrameworkBenefits} %\nameref{subsection:FrameworkBenefits} 
			    & \cref{fig:BenefitsDecision}  \\ 
			 RQ 3.2 (benefits and drawbacks) & team survey 
			    & \ref{subsection:FrameworkBenefits} %\nameref{subsection:FrameworkBenefits} 
			    & \cref{fig:BenefitsDecision}, \cref{fig:BenefitsOverview}\\
			 &   
			    & \ref{subsection:ResultsCorrelation} %\nameref{subsection:ResultsCorrelation}
			    & {\crefname{table}{tab.}{tab.}\cref{tab:Correlations}}  \\  
			 &   
			    &  \ref{subsection:OpenQuestions} %\nameref{subsection:OpenQuestions} 
			    &  \\
			 RQ 3.3 (effort's justification) & team survey  
			    &  \ref{subsection:FrameworkBenefits} %\nameref{subsection:FrameworkBenefits} 
			    & \cref{fig:EffortJustified}  \\  
			 
			 \midrule
			 RQ 4.1 (managers' TD overview) & management survey  
			    & \ref{subsection:managementSurvey} %\nameref{subsection:managementSurvey}  
			    &   \\ 
			 RQ 4.2 (benefits of the managers' TD overview)  & management survey 
			    & \ref{subsection:managementSurvey} %\nameref{subsection:managementSurvey}  
			    &   \\
		
			\bottomrule
			\end{tabular}
			\captionsetup{justification=centering}
		\caption{Overview of methods and figures used to answer the respective questions and results section containing details}
		\label{tab:ResultsOverview}
		\end{table*}
	
\begin{figure*} [t]
 		\centering
		\includegraphics[width=\linewidth]{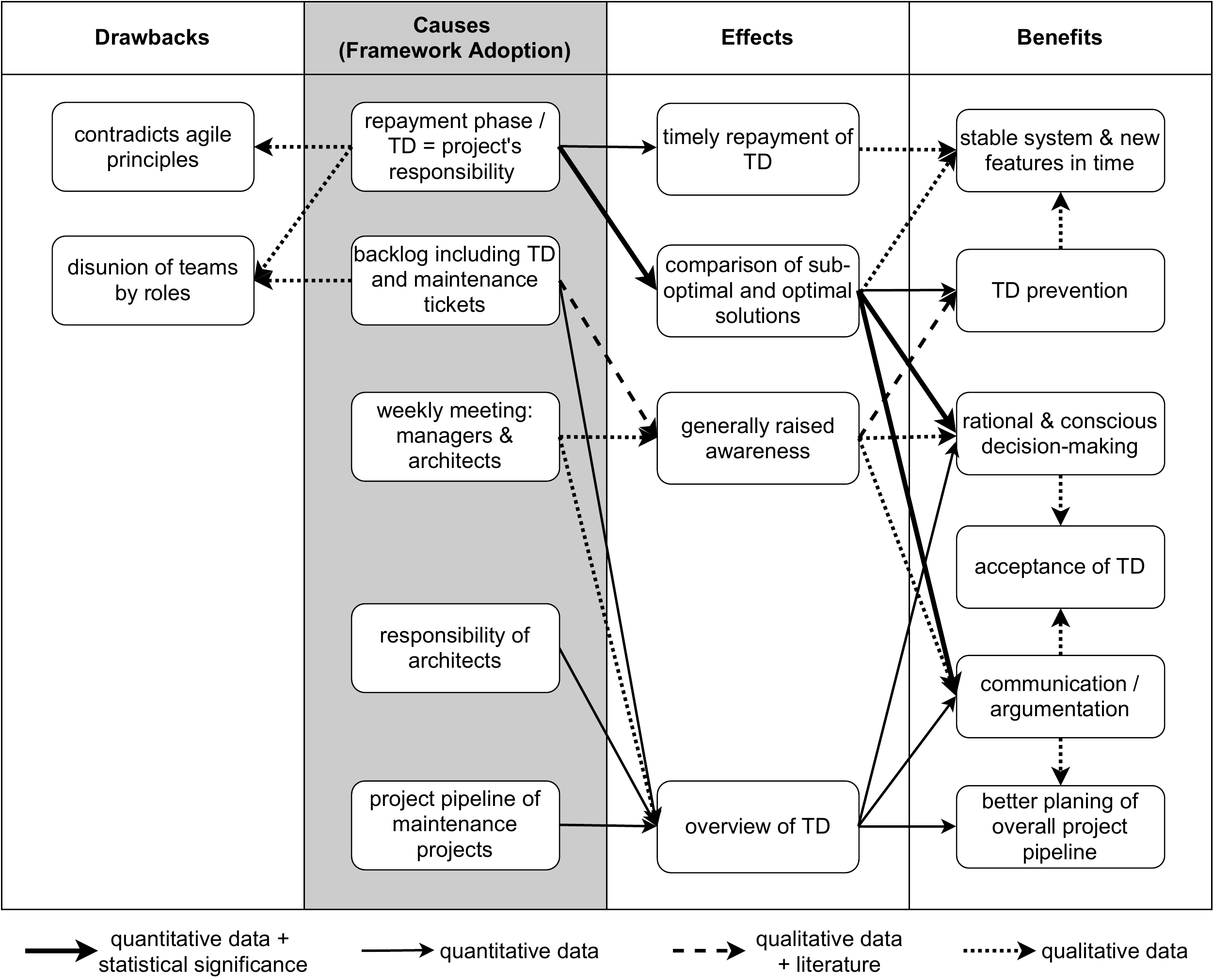}
		\caption{Causes, perceived effects, and perceived benefits and drawbacks (arrows show effect-direction) of the TAP framework}
 		\label{fig:EffectDiagram}
  	%	\Description{After the main deployment a phase of TD repayment follows.}
	\end{figure*}

    To give a better overview of the results, we generated a variation of a cause-effect diagram (\Cref{fig:EffectDiagram}) summarizing the effect, benefits, and drawbacks. 
    %\footnote{We did not use an Ishikawa diagram \cite{Best2008} because we wanted to present multiple outputs while an Ishikawa diagram focuses on one output.}
    The diagram is based on the qualitative and quantitative data found in this study.
    Moreover, we took logical deduction into account for some of the more detailed connections, e.g the repayment phase leads to timely repayment.
    The literature review led to some connections that supports the qualitative data.
    In addition, we summarized the codes at a higher level to give a better overview.
    %One may argue that some arrows are missing in this figure, e.g. ``rational \& conscious decision-making'' may lead to ``TD prevention''. 
    %This may be true but was not examined by our research. Therefore, we cannot make this declaration.
    The following sections discuss these cause-effect mechanisms in detail and answer the RQs.
    
    \subsection  {Application of the TAP Framework}
		\label{subsection:disRq}
		
	%	The basis of this work and the respective observations is the framework for managing TD as described in \Cref{section:framework}. 
	%	The RQ's are dedicated to feasibility, effects and benefits of this framework.

		\textit{\textbf{RQ 1.1/RQ 1.2:} Do practitioners find the TAP framework reasonable?
		Are the processes of the TAP framework feasible in practice?}
		
		\Cref{fig:RecordingRes,fig:RecordingProcRes,fig:RecordingProcWork,fig:RecordingRes,fig:ProcessingRes,fig:ProcessingProcRes,fig:ProcessingProcWork} shows little disagreement. Most team survey participants found the framework reasonable and thought it did work as intended. 
		Only a few participants thought it must be optimized.
		The ticket statistics (see \Cref{subsection:TicketStat}) indicate that the tickets were recorded and processed as intended.
		%Furthermore, the IT unit had already used the framework for more than two years.
		We conclude that the framework is feasible in practice and the following findings based on this framework are valid.
		
    \subsection {Perceived Effects of the TAP Framework}
        \textit{\textbf{RQ 2.1:} Is the use of the TAP framework associated with a raised awareness for the incurrence of TD?}
        
        %\Cref{fig:RecTD} shows most 
        On the one hand, 75\% of the participants stated they were aware of taking on TD at the point in time they incur it.
        There was no difference between the observed and the comparison unit.
        On the other hand, the answers to the open questions support the finding that the awareness was raised considerably (``TD is a topic'') and that the participants started to consider when to incur or keep TD and when not to (``Sometimes TD are OK'').
        In the management survey, the managers also imply that the awareness was raised. 
        This might indicate that the awareness before introducing the framework was lower in the observed unit than in the comparison unit. 
        However, we found out that the survey question on whether the participants are aware of the TD incurrence had great potential to be misunderstood. 
        Furthermore, developers may be biased to think they always know when they incur TD. 
        %This is also why we did not provide a figure for this result. 
        
        \textit{\textbf{RQ 2.2:} Is the use of the TAP framework associated with a more conscious incurrence of TD items?}
        
        \Cref{fig:CompareQvO} reveals that the observed unit showed a significantly different behavior than the comparison unit (\Cref{tab:MannWhitney}). 
        The observed unit's members compared different solutions and their costs before deciding on one of them.
        They may still have incurred TD by choosing the sub-optimal solution, but the benefit was a consciously and intentionally made decision.
        
        \textit{\textbf{RQ 2.3:} Is the use of the TAP framework associated with a better overview of TD items?} %--- here ---
        
        In \Cref{fig:WhoHasOverview}, it is striking that in the observed IT unit, the architects were the ones who had the overview of all TD items (100\% agreement).  
        The responsibility to keep an overview of all tasks rested with the architects.
        We could not observe such a clear responsibility in the comparison unit.
        %In the management survey, the managers stated that the weekly meetings with the architects resulted in them having a complete overview of their teams' TD. 
        A TD overview might also be created in the comparison unit by other means.
        However, the evaluation indicates that the clear responsibility of the architects, which led to the TD overview, was an additional benefit of the framework.
        
        \textit{\textbf{RQ 2.4:} Is the use of the TAP framework associated with a timelier repayment of TD items?}
        
        If the unit's members had to incur TD, they mostly paid it back timely, as seen in \Cref{fig:TDTbyTime}.
        %Even though not all TD tickets may be paid back during their respective projects as intended by the framework. 
        Additionally, \Cref{fig:BenefitsDecision} shows that two-thirds of participants in the observed IT unit perceived they repaid TD timelier. 
		Nevertheless, the unit did not repay all TD tickets during their respective projects, and one-third of participants were still dissatisfied with the timely repayment.
		
% 		\textit{\textbf{RQ4.2:} What are the effects of the overview of the TD from a management perspective?}
%         All IT managers agree that TD items become visible and thereby the overall awareness for TD increases.
%         The amount of TD can be used as a measure for the importance of projects.
%         By generating the comprehensive project pipeline the effects of the repayment become visible to the customers.  
        
        \textit{\textbf{RQ 4.1:} Do the managers have an overview of their IT systems' TD?}
        
        The observed unit's managers had an overview of their TD. 
        The architects were tasked to a) manage the maintenance project pipeline and b) track TD and other maintenance tasks in the backlog.
        The managers got their overview primarily due to the weekly meeting with the architects.
        They profited from the architectural evaluation of these tasks.
        The comparison unit's managers did not have a comprehensive overview of their TD.
        
    \subsection{Perceived Benefits and Drawbacks of the TAP Framework}
        \textit{\textbf{RQ 3.1/RQ 3.2:} Do practitioners observe that TD can be prevented by the adoption of the TAP framework?
        Are there other benefits or drawbacks arising from the adoption of the TAP framework?}
        
        The most apparent perceived effect of the framework was the conscious incurrence of TD by comparing different solution options.
        In \Cref{fig:BenefitsDecision}, we presented the benefits associated with comparing solution options.
        
        To back up these findings, we evaluated the correlation between the perceived benefits and the comparison of solution options and the estimations.
        We assume that the comparison and estimations can be associated with more rational discussions and decisions following the significant correlations.
        %The error probability for these assumptions was lower than 5\%. 
        %Therefore, we attribute these benefits to the application of the TAP framework.
        
        The correlations (\Cref{tab:Correlations}) do not support associations for TD prevention or the estimation of interest costs.
		The missing correlations for interest costs may be since only one-third of the participants estimate these costs.
		The missing correlations for TD prevention lead to the assumption that the prevention of TD is maybe not directly caused by the comparison and estimations.
		Still, TD prevention was perceived as a benefit by 82.4\% of the participants, and it was mentioned in the open questions. 
		Furthermore, it logically follows that rational decision-making may be able to prevent TD.
		
        The TD overview's perceived benefits were more rational decisions and better planning of the project pipeline by justifying TD repayments and integrating TD repayments into projects (\Cref{fig:BenefitsOverview}).
        
        Finally, the open questions support that the awareness for TD was raised, and activities like prioritization were discussed throughout the unit. Furthermore, optimized understanding and discussions were mentioned.
        %This awareness may enhance the prevention of TD as pointed out by the manager of the compared unit.
        
        The researchers observed no particular drawbacks other than the (initial) effort to use the framework (see RQ 3.3).
        Therefore, we asked for drawbacks in open questions.
        One participant each mentioned the lack of adherence to agile processes and the possible separating affect the framework may have on the team. 
        Finally, it was suggested to better involve the operations team.
        
        \textit{\textbf{RQ 3.3:} Do practitioners find these benefits justify the additional effort?}
        
		\Cref{fig:EffortJustified} shows that most participants of the team survey think that the perceived benefits justified the additional effort for comparison, cost estimation, and the generation of the overview.
		These results confirm the TAP framework's usefulness.
        
        \textit{\textbf{RQ 4.2:} What are the benefits of the TD overview from a management perspective?}
        %Awareness
        All IT managers agreed that TD items became visible and, thereby, the overall awareness for TD increased.
        %Project Pipeline
        In particular, the project pipeline's management profited from the visibility and architectural evaluation of TD. 
        The amount of TD could be used to measure the importance of projects.
        The project's durations became discernible, and the managers could postpone subsequent projects if necessary.
        %Prevention
        The managers noticed that decisions regarding the incurrence of TD were made more conscious and, by this, TD could be prevented. 
        %Communication
        The communication between IT management and teams improved.
        The better overview led to better communication with the customers and a better acceptance of repayment measures.
		Ultimately, TD management has been in the customer's interest as it provides stable systems and new features in time.

		\subsection  {Threats to Validity}
		\label{subsection:disTtV}

		\textit{\textbf{Construct Validity.}}
		To enhance the construct validity, we questioned and compared two units, one of which did adopt the TAP framework and one did not use any method for managing TD.
		As presented in \Cref{section:CaseDescription}, the scope and size of the comparison unit differ from the observed unit. 
		However, the organizational form, particularly the agile-managed projects with deadlines, is identical. 
		Since the time limitation is the most important basis for a meaningful framework adoption, we assume that the comparison of units is valid.
		
		To ensure comprehensibility and to optimize the questionnaire, we did pilot tests. 
		It could be a problem that the team survey did not ask directly but only in open questions for drawbacks. 
        Furthermore, the quantitative survey's construction was not preceded by a qualitative survey but only by the researcher's observations.
		The open questions and the question about the effort's justification should have addressed parts of this problem. 
		Additionally, we not only evaluated the questionnaire but also the ticket's statistics.
        This data triangulation improves the construct validity.
        For observer triangulation, two researchers independently coded the open questions of the team survey and the management survey.
	    %In a qualitative survey, we questioned the perspective of the managers.
	    Only three managers answered the follow-up management survey.
        Thus, this survey only provides qualitative insights.
        However, the unit manager who answered this survey is the co-author of this paper and may be biased. 
        Yet, this co-author was only involved in developing the framework and writing the framework section.
        He was not involved in the scientific evaluation. 
        Still, he may be biased toward the positive impact of the framework.
        Two co-authors utterly uninvolved with the company enhance the validity despite a possible author's bias.

	    %Most insights are found multiple times in literature and therefore provide more evidence to the body of knowledge. 
	    %The effects on the planning of project pipelines is a new and interesting aspect and should be evaluated in further research.
	
		\textit{\textbf{Conclusion Validity.}}
		The paper was able to show statistically significant differences between the observed and the comparison unit. 
		We used SPSS and standard statistic techniques to evaluate the findings' significance. 
		Nonetheless, the sample was very small as it can only encompass the two IT units, and the staff survey had an overall response rate of 55\%.
		We carefully considered the small sample size when interpreting the findings. 
		We suggest supporting the findings and assumptions by further replicating this study to gain more data points. 
		%The findings should, therefore, be taken as first hints on the benefits this framework might bring.
		%The findings should be supported by further replications to gain more data points.
		%Additionally, the comparison unit was considerably smaller than the observed unit and had a slightly different organization. 
		%We minimized this threat by choosing a comparison unit led by the same unit manager. 
		
		\textit{\textbf{Internal Validity.}}
        We observed valuable effects, but the same effects could have been observed in other IT units. 
		This threat was minimized by the comparison with a unit not using the TAP framework. 
		Nevertheless, just the management of TD could have led to better developer morale as presented by Besker et al. \cite{Besker2020c} and consequently to good ratings. 
		This threat can and should be minimized by replicating the study in other constellations. 
		As correlations do not directly lead to causalities, the mentioned interpretations of the correlations should be validated, e.g., by follow-up interviews.

		\textit{\textbf{External Validity.}}
	    The TAP framework has been developed and tested in an industrial environment, and it has been in use for a long time.
	    Both these facts are meaningful benefits of this research.
		Nevertheless, this was just one case study, and it may be biased. 
		We suggest replicating the framework adoption and the study in other units and companies, economic sectors, and countries to confirm the findings.
		Furthermore, the comparison unit should adopt the framework, or the reasons for not adopting this should be evaluated.
		We suggest focusing on the most relevant parts which were already mentioned in \Cref{section:FramworkScope}: the distinction in processing intentional TD as TD tickets and unintentional TD as part of maintenance tickets, the immediate recording of intentional TD, and the processing of intentional TD as part of the project they were incurred in. 
		
		For this, we provided the framework information in this paper and the questionnaires and raw data online\footnote{\url{https://doi.org/10.5281/zenodo.5788222}}. 
		%This study can be replicated using the provided information in this paper.
		%The data for replication and verification can be downloaded online\footnote{\url{https://doi.org/10.5281/zenodo.4616485}}. 

	\section{CONCLUSION AND FUTURE WORK}
	\label{section:conclusion}

         %\subsection  {Conclusion}
    	In this paper, we present and evaluate the TAP framework developed and used in practice for a long time.
    	%, the TAP framework %shows a high external validity and 
    	%provides the research community with relevant insights into the practical handling of TD.
    	The framework uses TD tickets for intentionally incurred TD to include TD management in the project management timeline. 
        These TD tickets transfer the responsibility for TD incurred during a project to the same project's project management.
        The team records TD tickets at the time they incur the TD.
        This creates a feedback loop of team and project management behavior in the observed unit. They perceived the following benefits:
    	%In the case company, the adoption of the framework causes a higher TD awareness
    	\begin{itemize}
            \item The communication between different stakeholders, including IT management and customers, is optimized, and discussions between the unit members are led more rationally.
            \item The overall awareness for TD rises by  making  TD visible and providing an overview of TD.
            %\item Project managers get an extrinsic motivation to reduce the incurrence of TD.
            \item TD tickets raise the awareness for the decision of whether TD should be incurred and make these decisions conscious and intentional. 
            \item TD items unintentionally incurred due to unconscious decisions can be prevented.
            \item TD items intentionally incurred during a project are part of the project and are paid back timelier.
        \end{itemize}
        
        Furthermore, TD tickets provide an in-time overview for the IT managers as these items are tracked during the project. 
        The management survey indicates that this enables the managers to react and adjust the project pipeline if necessary.

    	While the idea of maintenance tickets and projects is not new, the idea of TD tickets for intentional TD and the proposed processing of these tickets as part of the project is a novel approach.
    	To the best of the authors' knowledge, such an approach can not be found in the research literature so far.
    	
    	\textit{\textbf{For industry,}} this paper presents a framework that can be adopted and adapted to their own needs. 
    	The idea of including TD management in project management can be an incentive to develop a similar approach. 
    	The paper demonstrates the importance of raising TD awareness and of incurring TD intentionally by taking the time to compare different solution options. 
    	The advantages shown by this paper can be helpful when negotiating options to manage TD with IT management.
    	
    	\textit{\textbf{For researchers,}} this paper provides insights into procedures for managing TD developed in the industry. 
    	The paper indicates the importance of providing a solution for managing and preventing TD under the pressure of tight deadlines. 
    	Furthermore, this research confirms that the responsibility for TD needs to be transferred to the whole unit. 
        Developer teams cannot solve these problems alone. 
    	The TAP framework provides a solution for these problems. 
    	
    %	For future work, it can be worthwhile to evaluate if the comparison of solutions leads to better compromises that may not be the optimal solutions but one that may not need to be refactored either.
    %	Furthermore, the importance of raising the awareness for TD in the whole unit and not only the development team is an understudied area, and more ideas should be developed and evaluated.
    	\textit{\textbf{For future work,}} we believe it could be helpful to supplement the framework with more formal processes for some  TD activities, e.g., automated TD identification, TD measurement, or TD prioritization. 
    	In particular, ap\-proach\-es for TD prioritization may help optimize the repayment phase, e.g., a consequent distinction of potential and effective TD \cite{Schmid2013a}, as our research uncovered this to be a problem.
    	
    	The effects of TD management on the planning of project pipelines is an intriguing aspect. 
        Many underlying problems of TD management refer to communications issues between stakeholders, e.g., managers and developers. 
        Hence, we think that the perspectives of different stakeholders, e.g., line managers, on TD is another relevant research topic, which we would like to evaluate further.
        
    	At various points, we suggested replicating the study. 
    	However, this framework was designed as an actual frame to be adopted in whole or part.
		This is not a weakness of the framework, but takes into account the reality in which IT units' processes need to be adapted to their own needs. 
		We can see this, e.g., in the ITIL (Information Technology Infrastructure Library) processes or agile development methods.
		We hope to find other units that are willing to adopt all or some of the relevant parts of the framework mentioned in \Cref{section:FramworkScope}. 
		It would be interesting to evaluate which parts may lead to similar effects and benefits in other companies.

    	%Timely repayment of TD is sometimes not performed due to low priority or missing awareness of competing goals. Further research regarding project management and business value estimation is necessary. 
   % 	The timely repayment got negative feedback from some of the participants of the observed IT unit.
    %	It should be evaluated in what way this discontent could be managed, e.g. by prioritization or deferral.
	    %Finally, the prioritization of maintenance tasks including TD should be assessed between different application domains and industries.
	    
    	%One idea is to only repay high priority tickets during the project. %It is then another task to find a way to manage the deferred tickets. %, e.g. by linking the repayment to the next change of the affected part of the system.
    	%This will be part of our future work.

%%
%% The acknowledgments section is defined using the "acks" environment
%% (and NOT an unnumbered section). This ensures the proper
%% identification of the section in the article metadata, and the
%% consistent spelling of the heading.

\section*{Acknowledgment}
We would like to thank the observed and comparison unit of {\it Gruner+Jahr GmbH - Information Technology Department} for contributing to the framework, completing the questionnaires, and supporting this research.

%\section*{References}

\bibliography{TDAwareProjectmanagement}

\end{document}